\documentclass[10pt,twocolumn,letterpaper]{article}

\usepackage[pagenumbers]{wacv} %

\usepackage{latexsym}

\usepackage[accsupp]{axessibility}

\usepackage{multirow}
\usepackage{makecell}
\usepackage[finalizecache=false,frozencache=true,cachedir=minted-cache]{minted}
\newcommand{\mypara}[1]{\noindent\textbf{#1}}

\newcommand{\ourtitle}{SceneEval: Evaluating Semantic Coherence in Text-Conditioned\\3D Indoor Scene Synthesis}
\newcommand{\ours}{SceneEval\xspace}
\newcommand{\ourdataset}{SceneEval-500\xspace}

\definecolor{wacvblue}{rgb}{0.21,0.49,0.74}
\usepackage[pagebackref,breaklinks,colorlinks,allcolors=wacvblue]{hyperref}

\def\wacvPaperID{2328} %
\def\confName{WACV}
\def\confYear{2026}

\title{\ourtitle}

\author{
 Hou In Ivan Tam\textsuperscript{1},
 Hou In Derek Pun\textsuperscript{1},
 Austin T. Wang\textsuperscript{1},
 Angel X. Chang\textsuperscript{1,2},
 Manolis Savva\textsuperscript{1}
\\
 \textsuperscript{1}Simon Fraser University,
 \textsuperscript{2}Alberta Machine Intelligence Institute (Amii)
\\
  {\small{\url{https://3dlg-hcvc.github.io/SceneEval/}}}
}

\begin{document}

\maketitle
\begin{abstract}
Despite recent advances in text-conditioned 3D indoor scene generation, there remain gaps in the evaluation of these methods.
Existing metrics often measure realism by comparing generated scenes to a set of ground-truth scenes, but they overlook how well scenes follow the input text and capture implicit expectations of plausibility.
We present \ours, an evaluation framework designed to address these limitations.
\ours introduces fine-grained metrics for explicit user requirements---including object counts, attributes, and spatial relationships---and complementary metrics for implicit expectations such as support, collisions, and navigability.  
Together, these provide interpretable and comprehensive assessments of scene quality.
To ground evaluation, we curate \ourdataset, a benchmark of 500 text descriptions with detailed annotations of expected scene properties.
This dataset establishes a common reference for reproducible and systematic comparison across scene generation methods.
We evaluate six recent scene generation approaches using \ours and demonstrate its ability to provide detailed assessments of the generated scenes, highlighting strengths and areas for improvement across multiple dimensions.
Our results identify significant gaps in current methods, underscoring the need for further research toward practical and controllable scene synthesis.
\end{abstract}

\section{Introduction}
\label{sec:intro}

\begin{figure}
    \centering
    \includegraphics[width=\linewidth]{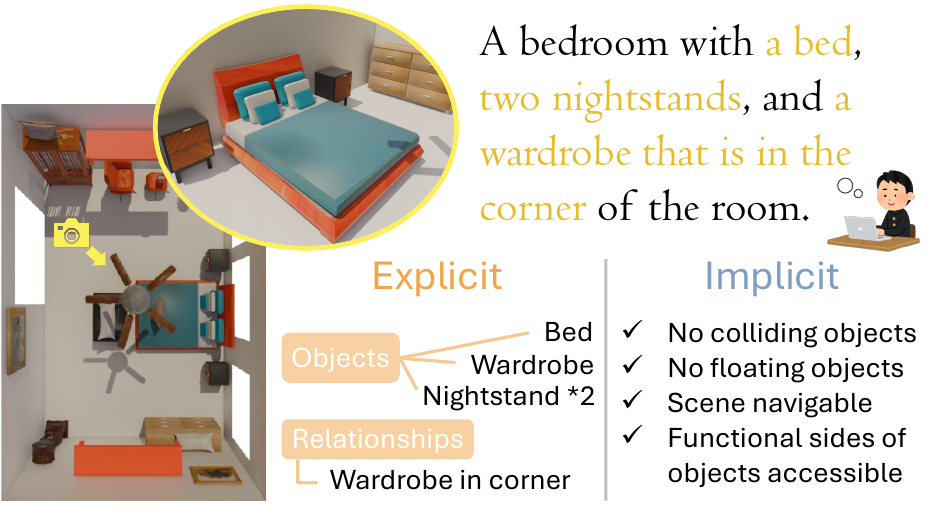}
    \vspace{-1.5em}
    \caption{
        Explicit vs. implicit requirements.
        Explicit requirements are communicated explicitly by the user in the text description, while implicit requirements are assumed but not necessarily stated.
    }
    \label{fig:explicit_implicit}
    \vspace{-1.75em}
\end{figure}

Digital 3D indoor scenes are essential for various applications, including robotics simulation, game development, and film production.
However, authoring 3D scenes manually is laborious, making automatic scene synthesis a long-standing research problem.
Scene synthesis faces two primary challenges: adhering to \textit{explicit} user requirements and meeting \textit{implicit} expectations, such as physical plausibility, which users often assume but do not explicitly specify.
As shown in \cref{fig:explicit_implicit}, both are crucial for practical applications.

Text-conditioned generation has been a popular research direction, allowing users to specify scenes through natural language.
Just as homeowners can convey their dream home to interior designers and scriptwriters can describe story scenes to set designers, natural language is an intuitive way for users to describe a desired scene to scene synthesis methods.
Depending on user expertise and needs, these descriptions can be minimalistic (``a cozy living room''), or more detailed (``a living room with a brown sofa facing a TV and a dining table with four chairs'').
The flexibility and expressiveness of text allows users to specify their desires without needing to understand the intricacies of 3D modeling.
This flexibility also poses unique challenges for scene synthesis methods, as they must understand and interpret the text descriptions to generate scenes that meet user requirements.
Users' unspoken expectations, such as object placements adhering to the laws of physics, further complicate this process.
Nonetheless, an ideal scene synthesis method should be able to generate scenes that satisfy both explicit and implicit user requirements.

Despite recent advances in scene generation methods, there is a lack of systematic evaluation of the \emph{fidelity} of the generated scenes against the input text descriptions.
Commonly used metrics fall into two categories.
\textit{Distributional metrics}, such as Fr\'echet Inception Distance (FID)~\cite{heusel2017gans} and categorical KL divergence, assess how realistic generated scenes are compared to a dataset of ground-truth scenes. 
However, this dependence on reference datasets makes them unsuitable for open-universe scene generation, where no ground truth exists.
\textit{Cross-modal metrics}, such as CLIP score~\cite{hessel2021clipscore}, measure text--scene alignment by computing similarity between rendered scene images and the input text descriptions, but they provide only a coarse sense of correspondence.
Neither type of metric reveals which specific constraints from the text are satisfied or violated, limiting insight into a method's strengths and weaknesses.
Beyond explicit fidelity to text, evaluation of implicit expectations is also incomplete, with prior work relying on isolated metrics such as collisions or out-of-bounds violations, which overlook broader aspects of physical plausibility (e.g., a scene may have no collisions yet remain unnavigable).
In contrast, text-conditioned generation in 2D modalities like image and video has seen more comprehensive evaluation. 
Yet 3D scene synthesis introduces unique challenges due to the added spatial dimension and physical constraints, making it non-trivial to adapt 2D evaluation metrics to 3D scenes and highlighting the need for a dedicated evaluation framework tailored to 3D scene synthesis.

To make systematic evaluation possible, there must be a shared benchmark that all methods can be evaluated against. 
However, prior work often uses ad hoc text descriptions and relies on qualitative inspection or user studies without annotated ground truth, hindering reproducible, fine-grained comparison.
To address this, we curate \emph{\ourdataset}, a collection of 500 indoor scene descriptions of varying complexity, each annotated with fine-grained ground-truth scene properties.
Each description is broken down into verifiable components, including object counts, attributes, spatial relationships, and architectural relations, providing a standardized reference for assessing explicit user requirements.

Building on \ourdataset, we introduce \emph{\ours}, a comprehensive framework that assesses both the satisfaction of explicit user requirements and the physical plausibility of generated scenes.
\ours defines four metrics for explicit constraints: \emph{object count}, \emph{object attributes}, \emph{object--object relationships}, and \emph{object--architecture relationships}, capturing the fine-grained details specified in the input text.
It further incorporates metrics for implicit expectations, including \emph{collisions}, \emph{support}, \emph{accessibility}, \emph{out-of-bounds}, and \emph{navigability}, offering a holistic view of physical plausibility.
By jointly evaluating explicit and implicit dimensions, \ours establishes a standardized benchmark that reveals the strengths and limitations of existing and future methods.

We evaluate six recent scene generation methods using \ours and demonstrate its effectiveness in providing better insights into their strengths and weaknesses.
Our results reveal that significant gaps remain in current approaches in generating scenes that fulfill explicit user requirements and meet implicit expectations, underscoring the need for further research in this area.
We believe \ours, together with \ourdataset, will serve as a valuable resource for developing methods that better align with user needs.
We will publicly release our code and dataset.

\section{Related Work}
\label{sec:related}

\begin{figure}
\includegraphics[width=\linewidth]{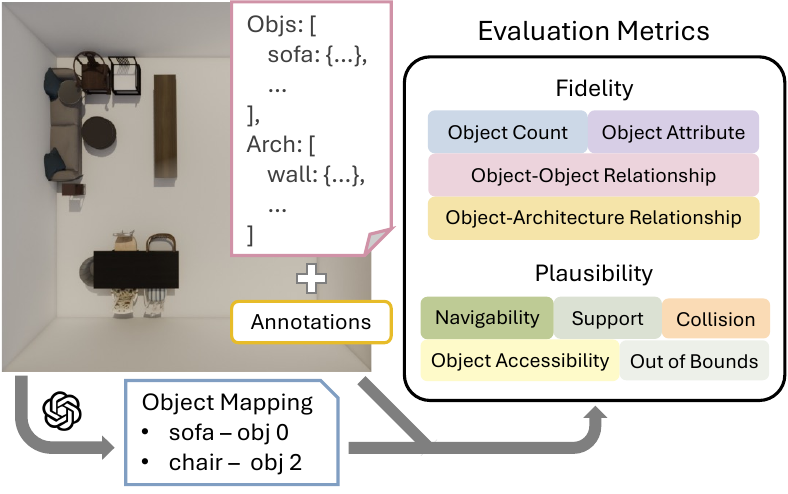}
\vspace{-2em}
\caption{
Overview.
Given a generated scene and its corresponding annotated properties, \ours first matches object instances in the scene to the annotated categories.
It then evaluates the scene on a comprehensive set of fidelity and plausibility metrics.
}
\vspace{-1.5em}
\label{fig:plausibility}
\end{figure}

\mypara{3D Indoor Scene Generation.}
Digital 3D indoor scene generation has been an active research area for decades.
Early work focused on developing systems to help users manually place objects in 3D scenes~\cite{bukowski1995object,shinya1995laying}.
Subsequent work has focused on automating generation with rule-based
~\cite{clay1996put,coyne2001wordseye,yeh2012synthesizing,deitke2022procthor,raistrick2024infinigen},
data-driven
~\cite{yu2011make,fisher2012example,chang2015text,savva2016pigraphs,ma2018language,keshavarzi2020scenegen}
and deep learning methods
~\cite{wang2018deep,li2019grains,ritchie2019fast,wang2019planit,paschalidou2021atiss,tang2024diffuscene,lin2024instructscene,zhai2024echoscene,zhai2024commonscenes,yang2024physcene,hu2024mixed,sun2025forest2seq,pfaff2025steerable,sun2025semlayoutdiff}.
These methods typically take a room type, floor plan shape, scene graph, or text description as input and aim to generate a 3D scene that satisfies the requirements.
In particular, text-conditioned generation has always been a popular direction for the appeal of specifying scenes with natural language.
With the advancement of large language models (LLMs) and vision language models (VLMs), many recent works
~\cite{aguina2024open,hu2024scenecraft,feng2024layoutgpt,ccelen2024design,yang2024holodeck,fu2024anyhome,wang2024architect,ling2025scenethesis,sun2025layoutvlm,gu2025artiscene}
incorporate them as both a text parser and a spatial prior for generating scenes with varying success.

\mypara{Evaluation of Text-conditioned Generation.}
Text-conditioned generation is widely studied across modalities such as text, images, video, and 3D shapes, each with metrics for measuring quality against input text.
In text generation, metrics like BLEU~\cite{papineni2002bleu}, ROUGE~\cite{lin2004rouge}, METEOR~\cite{banerjee2005meteor}, CIDEr~\cite{vedantam2015cider}, SPICE~\cite{anderson2016spice}, and BERTScore~\cite{zhang2019bertscore} assess alignment with reference text.
For images, CLIPScore~\cite{hessel2021clipscore}, BLIPScore~\cite{li2023blip}, and VQAScore~\cite{lin2025evaluating} evaluate text fidelity, while VBench~\cite{huang2024vbench}, VBench++~\cite{huang2024vbench++}, and WorldScore~\cite{duan2025worldscore} extend this to video and world generation.
In 3D shape generation, GPTEval3D~\cite{wu2024gpt} shows that GPT-4~\cite{achiam2023gpt} can assess text alignment and other aspects, while 3DGen-Bench~\cite{zhang20253dgen} builds on this using CLIP- and MLLM-based evaluators.
More recently, BlenderGym~\cite{gu2025blendergym} introduces VLM-based evaluation for scene editing, such as modifying object poses or materials.
However, these metrics do not transfer well to 3D scene generation.
Compared to 2D images, 3D scene descriptions are more ambiguous because of the added dimension.
For example, placing an object ``to the left'' of another can be interpreted differently depending on the viewpoint.
While GPTEval3D and 3DGen-Bench provide holistic evaluations, they lack the precision needed for fine-grained scene fidelity, offering only high-level insights.
These particularities, combined with the implicit expectations people have for plausible 3D environments, make evaluating text-conditioned 3D scene generation a unique challenge.

\mypara{Evaluation of Text-conditioned Scene Generation.}
Despite advances in scene generation, evaluation metrics have mostly focused on distributional similarity between real and generated scenes.
Common measures include Fréchet Inception Distance (FID)~\cite{heusel2017gans}, its CLIP-based variant FID\textsuperscript{CLIP}~\cite{kynkaanniemi2022role}, Kernel Inception Distance (KID)~\cite{binkowski2018demystifying}, scene classification accuracy (SCA), and categorical Kullback-Leibler divergence (CKL).
These metrics compare renderings or category distributions but do not evaluate scenes against input text, often necessitating user studies.
To address this gap, recent works~\cite{yang2024holodeck,hu2024scenecraft,wang2024architect,wang2024chat2layout,fu2024anyhome,ling2025scenethesis,gu2025artiscene,huang2025video} adopt CLIPScore and other text-image similarity metrics, while others~\cite{ccelen2024design,fu2024anyhome,wang2024architect} follow GPTEval3D~\cite{wu2024gpt} in using GPT-4~\cite{achiam2023gpt} or other MLLMs as evaluators.
These approaches measure overall correspondence between descriptions and renderings but provide little insight into which aspects of the text are preserved or lost.
Other directions involve annotating existing datasets with text descriptions and relationships~\cite{lin2024instructscene,ye2024relscene}, as well as scene-graph-based evaluations~\cite{zhai2024commonscenes,zhai2024echoscene}.
In contrast, \ours evaluates generated scenes against fine-grained properties specified in the input text, enabling a more detailed assessment of fidelity.
\ours further incorporates metrics for implicit expectations such as object accessibility and navigability, which are not the focus of prior work.

\section{\ours}
\label{sec:metrics}

Our goal is to evaluate how well a generated scene matches the user's request and whether it forms a physically plausible environment.
We capture these two dimensions as \emph{fidelity} (satisfying explicitly specified constraints such as object counts and attributes) and \emph{plausibility} (satisfying implicit expectations such as avoiding object collisions).
While many recent works rely heavily on VLMs as evaluators, we deliberately restrict their use to cases where they are truly necessary.
Whenever possible, we employ direct geometric checks, reducing reliance on VLM outputs and improving interpretability.
To enable such evaluation, we rely on annotations that describe the expected properties of a scene given its description, including object counts, attributes, and spatial relationships among objects and between objects and architecture.
Given a generated scene and its annotations, \ours first establishes a correspondence between the scene's objects and the annotated categories (\cref{sec:object_matching}), which then serves as the basis for computing our fidelity metrics (\cref{sec:text_fidelity_metrics}) and plausibility metrics (\cref{sec:plausibility}).
See the appendix for details on our dataset, implementation details of our metrics, and the VLM prompts.

\begin{table}
    \centering
    \resizebox{\linewidth}{!}
    {
    \begin{tabular}{@{} lrrrrrrr r@{}}
    \toprule
    Difficulty & Scenes & Words & Obj & ObjCount & ObjAttr & OORel & OARel \\ \midrule
    
    Easy    & 150 & 28.54 &  3.37 & 3.13 & 3.64 & 1.34 & 1.67 \\
    Medium  & 200 & 43.75 &  6.86 & 6.00 & 4.81 & 4.46 & 1.67 \\
    Hard    & 150 & 87.54 & 16.61 & 12.12 & 11.59 & 10.21 & 3.50 \\
    
    \bottomrule
    \end{tabular}
    }
    \vspace{-0.5em}
    \caption{
    Statistics of \ourdataset.
    Our dataset contains scene descriptions of three difficulty levels with increasing number of objects specified  (\textbf{Obj}).
    Each description is annotated with the expected properties in terms of object count (\textbf{ObjCount}), object attributes (\textbf{ObjAttr}), object-object relationships (\textbf{OORel}), and object-architecture relationships (\textbf{OARel}).
    }
    \label{tab:dataset_stat}
    \vspace{-0.5em}
\end{table}

\begin{figure}
    \centering
    \includegraphics[width=\linewidth]{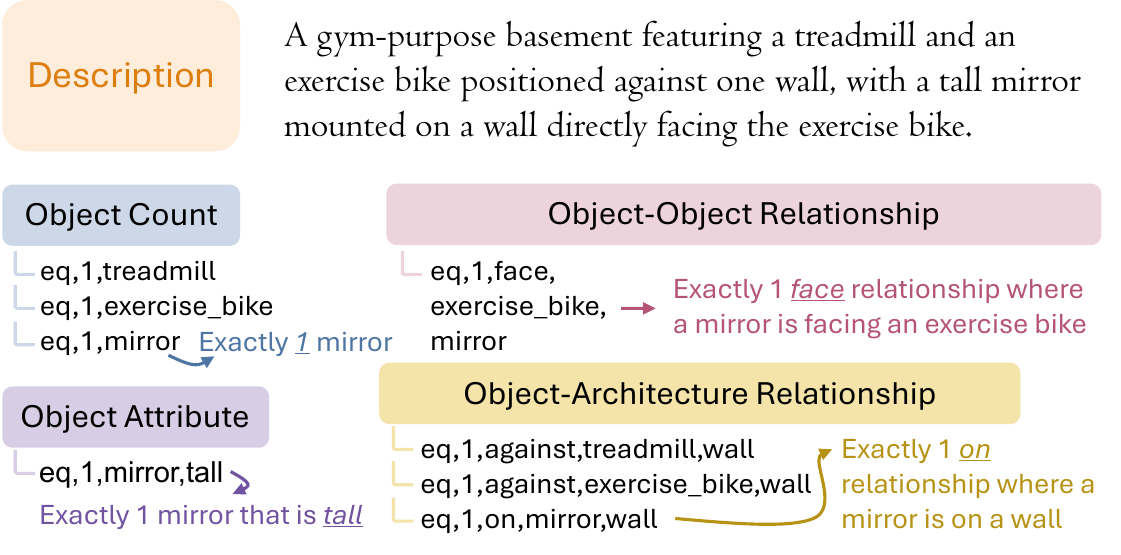}
    \vspace{-1.7em}
    \caption{
        Example entry of medium difficulty in \ourdataset.
        The scene description describes a basement room, a rarely-seen type in existing datasets.
        The annotation includes the expected scene properties, such as number of objects, specified in the text.
    }
    \label{fig:annotation}
    \vspace{-1.25em}
\end{figure}

\subsection{\ourdataset}

\begin{table*}
    \centering
    \resizebox{\linewidth}{!}
    {
        \begin{tabular}{@{} llll r@{}}
            \toprule
            Field & Schema & Example & Meaning \\ 
            \midrule

            Object Count &
            \texttt{quantifier}, \texttt{quantity}, \texttt{object category} &
            ge,2,bed &
            At least two beds. \\

            Object Attribute &
            \texttt{quantifier}, \texttt{quantity}, \texttt{object category}, \texttt{attribute} &
            eq,1,bed,king-size &
            Exactly one bed that is king-size. \\

            Object-Object Relationship &
            \texttt{quantifier}, \texttt{quantity}, \texttt{relationship}, \texttt{object category 1}, ... &
            eq,1,left,bed,lamp &
            Exactly one lamp to the left of a bed. \\

            Object-Architecture Relationship &
            \texttt{quantifier}, \texttt{quantity}, \texttt{relationship}, \texttt{object category}, \texttt{architecture type} &
            gt,2,against,bookshelf,wall &
            More than two bookshelves against a wall. \\

            \bottomrule
        \end{tabular}
    }
    \vspace{-0.5em}
    \caption{
        Annotation schema for \ourdataset. Each row shows the schema, an example annotation, and its natural-language interpretation. See \cref{sec:supp_annotation_schema} for more details on the schema.
    }
    \label{tab:annotation_schema}
    \vspace{-1em}
\end{table*}

\begin{table}
    \centering
    \resizebox{\linewidth}{!}
    {
    \begin{tabular}{@{} ll @{}}
    \toprule
    Manual & \makecell[cl]{
        A teenager's bedroom features a comfortable twin bed with a \\
        backboard in the far corner, with boxes underneath it. At the \\
        foot of the bed is a small desk equipped with a monitor, an \\
        external keyboard and mouse, and a desk lamp on the right for \\
        visibility, accompanied by a rolling chair. Next to the bed, a \\
        nightstand with an additional floor lamp nearby provides space \\
        for a phone and other valuables. A sizable wooden wardrobe with \\
        multiple drawers offers ample storage for clothes, while a coffee \\
        table beside it holds books and board games. In the center of the \\
        room, a tan-colored rug creates a cozy spot to sit, and the walls \\
        are adorned with various posters and pictures.
    } \\
    \midrule
    Generated & \makecell[cl]{
        A luxurious master bedroom features a canopy bed with an \\
        upholstered headboard. Two white nightstands stand beside \\
        it; one supports a marble table lamp and a photo frame, the \\
        other holds a green potted plant and an alarm clock. A tufted \\
        bench sits at the foot of the bed. Along one wall, a double-door \\
        wardrobe with patterned panels provides storage. Opposite, \\
        a wooden vanity desk topped with three metal candlesticks and \\
        a decorative tray is paired with an armless cushioned stool. A \\
        tall floor plant and a tripod floor lamp flank the desk on the \\
        left and right, respectively. Above a low dresser, a wall-mounted \\
        mirror reflects a crystal chandelier hanging from the ceiling. \\
        A plush area rug covers the floor under the bed.
    } \\

    \bottomrule
    \end{tabular}
    }
    \vspace{-0.5em}
    \caption{
        Two hard descriptions from \ourdataset, written by a human annotator (top) and generated by an LLM (bottom).
    }
    \label{tab:description_manual_llm}
    \vspace{-1.5em}
\end{table}

To evaluate how well generated scenes satisfy explicit requirements, we need structured annotations that translate natural language into machine-checkable constraints.
The Visual Genome dataset \cite{krishna2017visual} uses this idea in the 2D image domain by extracting scene graphs from images to capture objects and their relationships.
Analogously, we capture the key properties of 3D indoor scenes as described in text
 by providing the textual scene graph~\cite{schuster2015generating}.
To this end, we introduce \ourdataset, a dataset of 500 scene descriptions with annotations on the expected scene properties.
During annotation, we take each free-form description and provide a graph that specifies which objects should appear, what attributes they should have, and how they are arranged relative to one another and to the surrounding architecture.
This decomposition allows complex text descriptions to be checked component by component, providing a structured basis for evaluating scene quality.

We first constructed the dataset by manually writing 100 scene descriptions based on personal experience and reference images of homes and apartments.
Each description depicts an indoor environment with an emphasis on the objects it contains and their spatial relationships.
We did not impose a rigid template (e.g., always writing each sentence as ``object 1, relationship, object 2''), as our goal was to capture the diversity of how people naturally describe scenes.
We then carefully annotated each description according to our schema (\cref{tab:annotation_schema}), ensuring that the annotations faithfully reflect the content of the description.
In doing so, we paid special attention to nuances such as whether quantities are expressed exactly (e.g., ``there are two chairs'') or relatively (e.g., ``there are chairs,'' which implies more than one but not an exact number).
\cref{fig:annotation} shows an example entry from the dataset.
Together, the annotations form a structured representation of the described scene that can be automatically checked against a generated scene.

While this manual process ensures high quality, it does not scale to the size needed for systematic evaluation.
With recent advances in LLMs, we experimented with using an LLM (\textit{o4-mini-2025-04-16}~\cite{openai2025gpto4}) to scale up the dataset by generating additional scene descriptions together with their annotations.
We prompt the model with small batches of three to five descriptions at a time, using a set of manually written entries as in-context examples.
The model first produces new free-form descriptions and then fills in the corresponding annotations.
However, the raw LLM outputs often contain systematic errors, including missing annotations, hallucination of constraints from earlier generations, inconsistent formatting, and reduced diversity with longer generation history.
To address these issues, we carefully validate all generated entries: reading each description, rejecting ones that are too similar to existing entries, prompting the model for greater variation when needed, and correcting annotation errors by referencing the validated text.
See \cref{sec:supp_limitations_generation} for a discussion of the LLM's failure modes.
This process ensures consistency with the manually created annotations while preserving diversity in the dataset.
In practice, most automatically generated outputs required at least one manual edit, indicating that further work is needed for fully automatic generation.
Nevertheless, this semi-automatic approach substantially improves scalability, yielding 400 additional entries on top of the 100 manually created ones.
\cref{tab:description_manual_llm} presents examples of manually written and LLM-generated (and manually validated) descriptions and shows that they are comparable in quality.

Together, the manual and semi-automatic processes result in a total of 500 scene descriptions with annotations.
These cover ten common room types: bedroom, living room, dining room, playroom, gaming room, kitchen, bathroom, basement, den, and office.
We organize the dataset into three difficulty levels based on the complexity of the descriptions.
Easy descriptions specify at most four large furniture objects (e.g., bed, sofa).
Medium descriptions specify five to eight objects, with up to three being small objects (e.g., cup, book).
Hard descriptions contain at least nine objects, with no restriction on type, and may involve multiple rooms.
\cref{tab:dataset_stat} summarizes the dataset statistics.

\subsection{Object Matching}
\label{sec:object_matching}

Given a scene generated from a description in our dataset, we first need to establish reliable correspondences between the objects in the scene and the categories specified in the annotations.
This step is necessary because metadata of objects in generated scenes may be incomplete or unreliable, making direct comparison to annotations infeasible.
To address this, \ours renders a front-view image for each object and uses a VLM to check whether it matches the annotated categories.
Each annotated category can have zero or more corresponding instances in the scene, and categories without a match remain unmatched.
This mapping then serves as the basis for evaluating fidelity metrics.

\subsection{Text Fidelity Metrics}
\label{sec:text_fidelity_metrics}

Text fidelity is the extent to which a generated scene matches its input text description.
Because a description specifies objects, their attributes, and how they are arranged relative to one another and to the surrounding architecture, a comprehensive evaluation must consider all of these aspects.
To this end, \ours introduces four metrics: object count, object attribute, object--object relationship, and object--architecture relationship.
Each metric builds on the previous one, starting from the simplest requirement (having the right objects) and progressively addressing more detailed constraints.

\mypara{Object Count (CNT)}
is the most basic check: whether the number of objects in the scene matches the quantities specified in the description.  
For example, the text may require two chairs and one table.
Using the object mapping, we compare the instance counts in the scene to the annotated quantities and report the percentage of satisfied specifications.
This metric confirms that the building blocks of the scene are present, but it does not verify whether the objects look as intended.

\mypara{Object Attribute (ATR)}
extends the evaluation to whether objects have the correct attributes, such as a \emph{red} sofa or a \emph{wooden} table.
We render two images for each relevant object: one front view and one with a 170 cm human figure for scale.
These are provided to a VLM together with the annotated attributes, and the model judges whether the objects satisfy them.
This metric captures how well the scene reflects the descriptive details, but it still ignores how objects are placed relative to one another.

\mypara{Object--Object Relationship (OOR)}
addresses this gap by checking whether the placements of objects satisfy the spatial relationships specified in the description.
For example, a sofa next to a coffee table or a chair in front of a desk.
We define 13 types of spatial relationships, including \texttt{inside}, \texttt{side\_of}, and \texttt{next\_to} (see \cref{sec:supp_spatial_relationships}).
Because annotations are written in open vocabulary, we additionally map annotated relationships to these categories using a VLM (e.g., ``at the foot of a bed'' is mapped to \texttt{front\_of} and \texttt{next\_to}).
We then apply geometric techniques such as ray casting, point sampling, and position analysis to verify these relationships.
\Cref{sec:supp_implementation_details} provides further implementation details.
This metric evaluates relative positioning among objects, but many descriptions also tie objects to architectural elements.

\mypara{Object--Architecture Relationship (OAR)}
completes the set by checking relationships between objects and architectural elements.  
For instance, a sofa against a wall or a rug in the middle of a room.
We define 10 such relationships, including \texttt{against\_wall}, \texttt{corner\_room}, and \texttt{hang\_ceiling} (see \cref{sec:supp_spatial_relationships}).
The architectural reference can be a wall, floor, ceiling, door, window, or room.  
We use a similar process as in OOR to verify these relationships and report the percentage of satisfied specifications.  

Together, these four metrics provide complementary aspects of fidelity: object presence, object properties, spatial relations among objects, and spatial relations to the architecture.

\subsection{Plausibility Metrics}
\label{sec:plausibility}

Text descriptions specify what a scene should contain, but they often leave implicit many assumptions that humans naturally expect.
For example, a description rarely states that objects should not intersect, that they should be stably placed on surfaces, or that a person should be able to walk around the room.
Yet these assumptions are essential for scenes to be physically plausible and practically usable.
To capture them, \ours evaluates five aspects of plausibility: object collision, object support, scene navigability, object accessibility, and object out-of-bounds.
Together, these metrics move from basic physical feasibility toward more functional and spatial considerations.

\mypara{Object Collision (COL)}
is the most fundamental requirement: objects in a scene should not intersect with one another.  
We perform mesh-based collision tests between all pairs of objects and report the percentage of objects that are in collision.  
This ensures that scenes do not violate basic physical constraints, but it does not guarantee that objects are placed in a stable way.

\mypara{Object Support (SUP)}
addresses this by checking whether objects are stably supported.
We classify objects into four support types: \texttt{ground}, \texttt{object}, \texttt{wall}, and \texttt{ceiling}, by using a VLM to judge from rendered images.
Based on the assigned type, we apply ray casting to verify whether objects are supported by other objects or by architectural elements.
We report the percentage of objects that are correctly supported.
See \cref{sec:supp_implementation_details} for more details.
This metric ensures stability, but it does not address whether the overall arrangement allows people to move through the space.

\mypara{Scene Navigability (NAV)}
evaluates whether the arrangement of objects leaves sufficient connected free space for movement.
A scene with poor navigability may trap regions behind objects, making parts of the room unreachable.
Following PhyScene~\cite{yang2024physcene}, we define free space as the floor area not occupied by objects or architectural elements, and measure navigability as the ratio of the largest connected free space to the total free space.
We compute this by projecting the scene onto a 2D occupancy mask and applying connected component analysis.
While this evaluates movement at the room level, it does not ensure that individual objects remain usable.

\mypara{Object Accessibility (ACC)}
addresses this by checking whether the functional sides of objects are accessible.
Objects such as sofas, beds, and wardrobes have sides intended for interaction (e.g., the front of a sofa, the three sides of a bed, or the front of a wardrobe).
For each object, we use a VLM to identify functional sides based on its description, then apply a 2D occupancy analysis similar to NAV to check whether those sides are blocked.
We report accessibility as the ratio of unoccupied pixels to the total functional area, taking the best score if multiple sides exist.
This metric complements navigability by focusing on the usability of individual objects.

\mypara{Object Out-of-Bounds (OOB)}
ensures that objects remain inside the floor plan of the scene.
Without this check, a method could trivially satisfy the previous four metrics by placing objects outside the room.  
To prevent such cases, we sample points on each object's surface and cast rays toward the floor.
If fewer than 99\% of the points intersect the floor, the object is considered out-of-bounds.
This final metric enforces consistency with the room layout, complementing the checks provided by the other four metrics.

Together, these five metrics capture complementary aspects of plausibility: avoiding collisions, ensuring stable support, enabling movement, preserving object usability, and respecting room boundaries.

\begin{figure*}
    \centering
    \includegraphics[width=\linewidth]{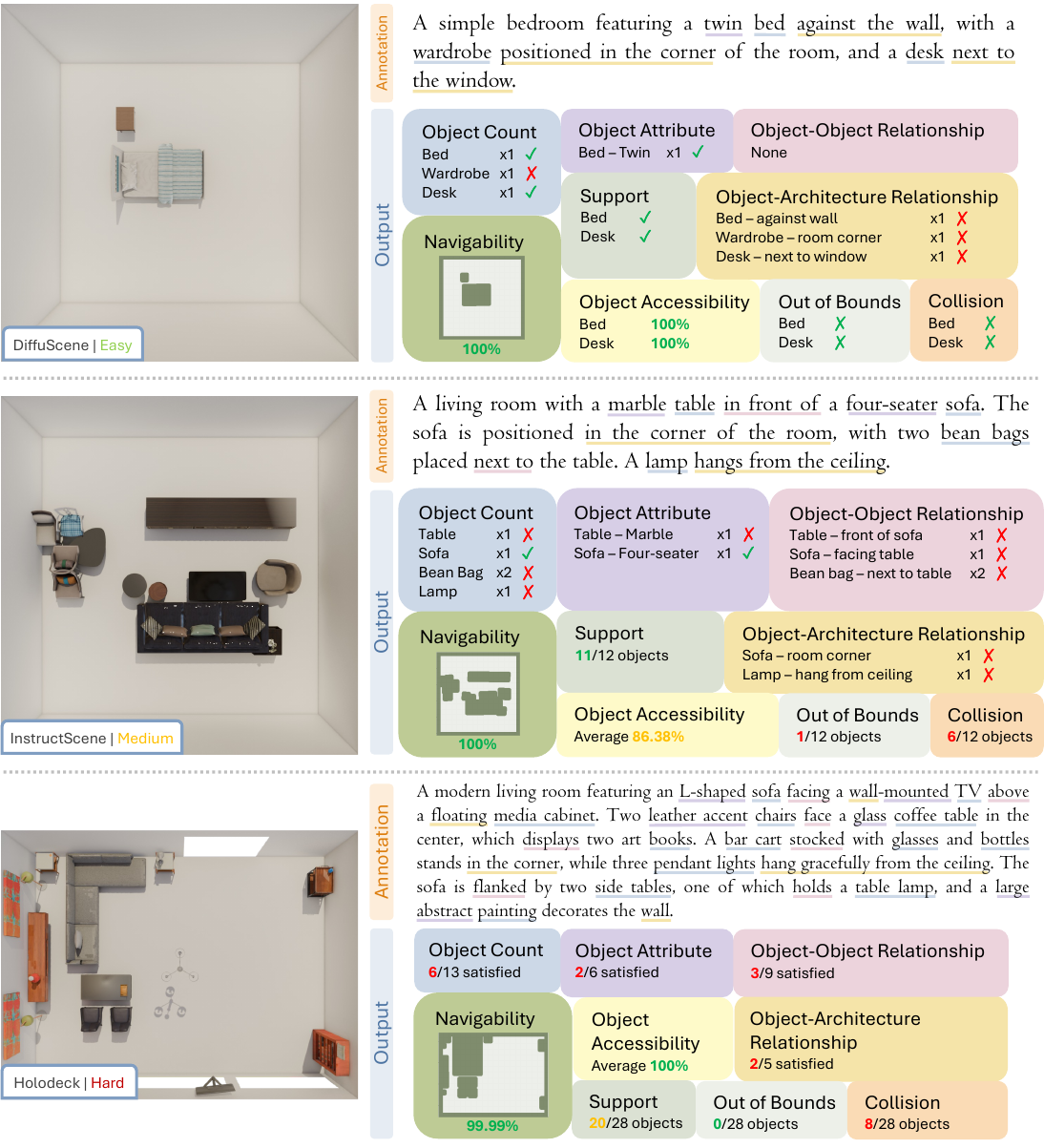}
    \vspace{-1.5em}
    \caption{
        Examples of scenes generated using text descriptions in \ourdataset and the corresponding evaluation results using \ours.
        Our dataset has scene descriptions with annotations of three difficulty levels: easy, medium, and hard.
        \ours provides a comprehensive evaluation of the generated scenes on fidelity and plausibility.
    }
    \label{fig:results}
    \vspace{-0.5em}
\end{figure*}

\section{Experiments}
\label{sec:results}

\begin{table*}
\centering
\resizebox{\linewidth}{!}
{
\begin{tabular}{@{} l rrrr rrrrrr rr r r@{}}
\toprule
& \multicolumn{4}{c}{Fidelity} & \multicolumn{6}{c}{Plausbility}
& \multicolumn{2}{c}{Resource}
& 
\\ 
  
\cmidrule(l){2-5} \cmidrule(l){6-11} \cmidrule(l){12-13}
& $\uparrow~$CNT$_{\%}$ & $\uparrow~$ATR$_{\%}$ & $\uparrow~$OOR$_{\%}$ & $\uparrow~$OAR$_{\%}$
& $\downarrow~$COL$_{ob\%}$ & $\downarrow~$COL$_{sc\%}$ & $\uparrow~$SUP$_{\%}$ & $\uparrow~$NAV$_{\%}$
& $\uparrow~$ACC$_{\%}$ & $\downarrow~$OOB$_{\%}$
& Mem$_{GB}$ & Time$_{sec}$
& CLIP\textsubscript{sim}
\\
\midrule

ATISS
& 11.18 & 7.40 & 1.07 & 8.03
& 50.36 & 75.20 & \textbf{90.90} & 99.83
& 85.55 & 10.86
& 0.15 & 1.52
& 15.69
\\
DiffuScene
& 11.99 & 9.28 & 3.20 & \textsuperscript{\textdagger}8.21
& 31.81 & 62.80 & 75.40 & \textsuperscript{\textdagger}99.44
& 81.86 & \textsuperscript{\textdagger}25.79
& 1.82 & 11.50
& 16.33
\\
LayoutGPT
& 11.84 & 8.05 & 1.18 & 4.87
& \textbf{11.46} & \textbf{27.20} & 30.13 & \textbf{99.99}
& 47.26 & 72.25
& - & 8.86
& 16.51
\\
InstructScene
& 14.14 & 11.53 & 3.59 & \textsuperscript{\textdagger}10.20
& 55.00 & 85.80 & 80.79 & \textsuperscript{\textdagger}98.99
& 77.47 & \textsuperscript{\textdagger}19.43
& 1.50 & 4.64
& 16.44
\\
LayoutVLM
& \textbf{35.59} & 20.20 & 6.03 & 19.39
& 32.13 & 57.80 & 76.90 & 99.65
& 85.19 & 4.89
& 3.76 & 89.09
& 15.83
\\
\midrule
Holodeck
& 32.64 & \textbf{28.49} & \textbf{11.52} & \textbf{37.27}
& 15.91 & 72.20 & 63.21 & 99.60
& \textbf{89.65} & \textbf{1.44}
& - & 98.87
& 17.63
\\

\bottomrule
\end{tabular}
}
\vspace{-0.5em}
\caption{
Evaluation of scene generation methods with \ours on our dataset.
We report averages across all scenes along with peak GPU memory usage and generation time.
\textsuperscript{\textdagger} indicates floor mask-based metrics where the method cannot be conditioned on it.
Holodeck achieves the strongest overall fidelity, though LayoutVLM slightly surpasses it on object counts.
While LayoutGPT appears to perform well on plausibility metrics, its high OOB suggests that it is not respecting the scene boundaries, showing the importance of a comprehensive evaluation.
We also compare against CLIP similarity between scene images and their text descriptions to show correlation with \ours.
}
\label{tab:result}
\end{table*}

\begin{table}
\centering
\resizebox{\linewidth}{!}
{
\begin{tabular}{@{}l l cccc c r@{}}
\toprule
&& \multicolumn{4}{c}{InstructScene} & LayoutGPT \\
\cmidrule(l){3-6} \cmidrule(l){7-7}
                            && $\downarrow$ FID & $\downarrow$ FID\textsuperscript{CLIP}& $\downarrow$ KID\textsubscript{$\times$1e-3} & SCA & $\downarrow$ FID \\
\midrule
\multirow{4}{*}{Bedroom}
& ATISS                      & 119.73 & 6.95 &  0.39 & 59.17 & 30.02 \\
& DiffuScene                 & 123.09 & 7.13 &  0.39 & 60.49 & -     \\
& LayoutGPT                  & -      & -    &  -    & -     & \textbf{29.88} \\
& InstructScene              & \textbf{114.78} & \textbf{6.65} &  \textbf{0.32} & \textbf{56.02} & -     \\
\midrule
\multirow{4}{*}{Living Room}
& ATISS                      & 117.67 & 6.08 & 17.60 & 69.38 & 85.40 \\
& DiffuScene                 & 122.20 & 6.10 & 16.49 & 72.92 & -     \\
& LayoutGPT                  & -      & -    &  -    & -     & \textbf{78.60} \\
& InstructScene              & \textbf{110.39} & \textbf{5.37} &  \textbf{8.16} & \textbf{65.42} & -     \\
\midrule
\multirow{4}{*}{Dining Room}
& ATISS                      & 137.10 & 8.49 & 23.60 & 67.61 & -     \\
& DiffuScene                 & 145.48 & 8.63 & 24.08 & 70.57 & -     \\
& LayoutGPT                  & -      & -    &  -    & -     & -     \\
& InstructScene              & \textbf{129.76} & \textbf{7.67} & \textbf{13.24} & \textbf{64.20} & -     \\
\bottomrule
\end{tabular}
}
\vspace{-0.5em}
\caption{
    Commonly reported metrics reproduced from InstructScene~\cite{lin2024instructscene} and LayoutGPT~\cite{feng2024layoutgpt}.
    These metrics provide a coarse sense of similarity to ground-truth scenes but do not reveal whether methods satisfy specific textual constraints or implicit plausibility requirements.
}
\label{tab:baseline}
\vspace{-2em}
\end{table}

\subsection{Evaluated Methods}
\label{sec:evaluated_methods}

We evaluate 
ATISS~\cite{paschalidou2021atiss}, DiffuScene~\cite{tang2024diffuscene}, InstructScene~\cite{lin2024instructscene}, LayoutGPT~\cite{feng2024layoutgpt}, LayoutVLM~\cite{sun2025layoutvlm}, and Holodeck~\cite{yang2024holodeck}.
ATISS is an early transformer-based model that generates indoor scenes conditioned on the room type and floor plan shape.
While it cannot be conditioned on text descriptions, it is often used as a baseline in recent work, and we include it to evaluate the importance of being able to condition on text descriptions.
All other methods we evaluate can be conditioned on text descriptions.
DiffuScene models scene generation as a diffusion process.
InstructScene incorporates a semantic scene graph as an intermediate representation and uses graph diffusion to generate scenes.
LayoutGPT is a pioneering work that uses LLM to generate indoor scenes as CSS code.
Following the trend of using VLMs for world knowledge, Holodeck is an extensive system designed to generate multi-room indoor scenes for embodied AI simulation, and LayoutVLM is a recent method that incorporates differentiable optimization into a VLM-based framework for generating scene layouts.

ATISS, DiffuScene, and InstructScene are all trained on the 3D-FRONT dataset~\cite{fu20213dfront, fu20213dfuture}.
While LayoutGPT does not involve training, it  provides scenes from 3D-FRONT as in-context examples to the LLM.
Holodeck and LayoutVLM do not require a pre-existing scene dataset and use assets from Objaverse~\cite{deitke2023objaverse}.
All methods use retrieval to obtain 3D assets for the scenes, and we use the same asset sources as in the original work.
DiffuScene and InstructScene cannot be conditioned on a floor plan shape, while the others can.
Only Holodeck can generate architectural elements (floors, walls, windows, doors, and ceilings) in addition to objects.
For other methods, we provide a 6 m $\times$ 6 m square floor plan shape with walls as input.
Generations requiring a GPU are run on a single NVIDIA RTX 4090 with 24 GB of VRAM, totaling less than 2 hours of GPU time across all methods.
For all VLM usage, we use \textit{GPT-4o-2024-08-06}~\cite{achiam2023gpt} with the default parameters. 
Note that while we choose to evaluate these six methods, \ours is a general evaluation framework applicable to a wide range of scene generation methods.

\subsection{Analysis}
\label{sec:analysis}

\cref{fig:results} shows several generated scenes from the methods given text descriptions from \ourdataset and the corresponding evaluation.
\cref{tab:result} presents the overall quantitative results.
We also report the average peak GPU memory usage and generation time per scene, with a breakdown by description difficulty in \cref{tab:result_by_difficulty} in the appendix.

To assess whether \ours aligns with human judgment, we manually evaluated 500 scenes generated from the 100 manual descriptions for fidelity.
We found agreement rates of 89.8\%, 83.5\%, 94.6\%, and 94.1\%, and Cohen's kappas of 0.75, 0.56, 0.72, and 0.77 for CNT, ATR, OOR, and OAR, respectively.
We also conducted a user study with 10 participants on 125 scenes, yielding agreements of 89.5\%, 89.9\%, 91.7\%, and 88.1\%, with Cohen's kappas of 0.72, 0.57, 0.58, and 0.56. 
These show that \ours is consistent with human judgment.
See the appendix for the user study instructions.

\textbf{SceneEval provides interpretable evaluation.}
Existing distributional and cross-modal metrics such as FID, KID, SCA, or CLIP\textsubscript{sim} (\cref{tab:baseline}, last column in \cref{tab:result}, computed with Long-CLIP-L~\cite{zhang2024longclip}) allow relative comparison between methods, but they provide little insight into \emph{why} one method is better than another.
For example, the higher FID or lower CLIP\textsubscript{sim} of DiffuScene does not illuminate whether this is due to poor scene layout, unrealistic object appearances, or other visual inconsistencies.
Similarly, Holodeck's +1.2 improvement in CLIP\textsubscript{sim} over InstructScene is hard to interpret---does it reflect more accurate object attributes, fewer collisions, or simply rendering differences?
These image-based metrics are sensitive to rendering, making them unreliable when comparing methods that use different asset sources.
In addition, reliance on specific scene datasets for evaluation also renders prior metrics inapplicable to models that do not have a reference dataset, which is often the case for newer VLM-based methods.
In contrast, \ours evaluates scenes directly against input text descriptions, disentangling object counts, attributes, relationships, and plausibility.
This provides interpretable feedback on which constraints are satisfied and which are violated, offering deeper diagnostic value across a broader range of methods.
This interpretability advantage becomes even more important when evaluating plausibility, where single metrics can be misleading.

\textbf{Scene plausibility requires comprehensive evaluation.}  
A scene may appear plausible under one metric while failing badly under another, which makes it essential to evaluate multiple aspects together.
For example, LayoutGPT achieves the best scores in object collision and scene navigability, suggesting its scenes are both physically consistent and easy to move through.
However, it simultaneously has the worst out-of-bounds and support rates, revealing that many objects are either placed outside the room or without stable support.
By doing so, LayoutGPT can trivially reduce collisions and free up navigable space, giving the illusion of plausibility.
This case illustrates why partial evaluation is misleading: without OOB or support checks, LayoutGPT would have been judged the most plausible method.
In contrast, \ours exposes such failure modes by combining complementary metrics that together provide a complete view of scene plausibility.

\textbf{Holodeck has the best overall fidelity.}
Across the four fidelity metrics, Holodeck achieves the strongest overall performance, though LayoutVLM slightly surpasses it on object counts.
In contrast, several other methods struggle even with simple constraints such as producing the correct number of objects.
A common factor among these weaker methods is their reliance on the 3D-FRONT dataset, whose limited diversity may contribute to poor generalization.
While 3D-FRONT has been a valuable resource for training scene generation models, these results suggest that future work should examine the effectiveness of this dataset in creating methods that align with actual user needs.

\textbf{Limits of fine-grained fidelity.}
All methods, including Holodeck, struggle when it comes to fine-grained constraints.
Even the best method satisfies fewer than 30\% of attribute requirements.
We hypothesize that this reflects limitations in retrieval mechanisms, which often fail to retrieve objects with the right attributes, compounded by limited diversity in the underlying datasets.
As a result, descriptive details such as color, material, or style are not well captured.
Performance is even weaker on object-object relationships: no method satisfies more than 20\% of the specified constraints.
At this rate, users have little to no control over how objects are placed with respect to each other, which is a critical limitation for practical applications.
Together, these results show that while current models can capture object categories and sometimes counts, they fail to translate richer textual specifications into scene structure.
Closing this gap is an important direction for future research toward practical scene synthesis.

\section{Conclusion}
\label{sec:conclusion}

We presented \ours, a comprehensive framework for evaluating text-conditioned 3D indoor scene generation.
By defining metrics that separately assess explicit user requirements and implicit expectations, \ours enables more interpretable and diagnostic evaluation than commonly used metrics.
To support systematic evaluation, we curated \ourdataset, a benchmark of 500 scene descriptions with fine-grained annotations, which establishes a shared reference point for future research.
Our experiments on six recent methods revealed consistent shortcomings in the methods' ability to satisfy detailed user constraints and ensure physical plausibility.
These limitations highlight the importance of evaluation frameworks that reveal the true capabilities of existing methods.
We believe \ours, together with \ourdataset, will be a valuable tool for building scene generation methods that better align with user needs and expectations.

\section*{Acknowledgments}
\label{sec:ack}

This work was funded in part by the Sony Research Award Program, a CIFAR AI Chair, a Canada Research Chair, NSERC Discovery Grants, and enabled by support from the \href{https://alliancecan.ca/}{Digital Research Alliance of Canada}.
We thank Nao Yamato, Yotaro Shimose, and other members on the Sony team for their feedback.
We also thank Qirui Wu, Xiaohao Sun, and Han-Hung Lee for helpful discussions.

{
    \small
    \bibliographystyle{ieeenat_fullname}
    \bibliography{main}

\begin{thebibliography}{67}
\providecommand{\natexlab}[1]{#1}
\providecommand{\url}[1]{\texttt{#1}}
\expandafter\ifx\csname urlstyle\endcsname\relax
  \providecommand{\doi}[1]{doi: #1}\else
  \providecommand{\doi}{doi: \begingroup \urlstyle{rm}\Url}\fi

\bibitem[Achiam et~al.(2023)Achiam, Adler, Agarwal, Ahmad, Akkaya, Aleman, Almeida, Altenschmidt, Altman, Anadkat, et~al.]{achiam2023gpt}
Josh Achiam, Steven Adler, Sandhini Agarwal, Lama Ahmad, Ilge Akkaya, Florencia~Leoni Aleman, Diogo Almeida, Janko Altenschmidt, Sam Altman, Shyamal Anadkat, et~al.
\newblock {GPT-4} technical report.
\newblock \emph{arXiv preprint arXiv:2303.08774}, 2023.

\bibitem[Aguina-Kang et~al.(2024)Aguina-Kang, Gumin, Han, Morris, Yoo, Ganeshan, Jones, Wei, Fu, and Ritchie]{aguina2024open}
Rio Aguina-Kang, Maxim Gumin, Do~Heon Han, Stewart Morris, Seung~Jean Yoo, Aditya Ganeshan, R~Kenny Jones, Qiuhong~Anna Wei, Kailiang Fu, and Daniel Ritchie.
\newblock Open-universe indoor scene generation using {LLM} program synthesis and uncurated object databases.
\newblock \emph{arXiv preprint arXiv:2403.09675}, 2024.

\bibitem[Anderson et~al.(2016)Anderson, Fernando, Johnson, and Gould]{anderson2016spice}
Peter Anderson, Basura Fernando, Mark Johnson, and Stephen Gould.
\newblock {SPICE}: Semantic propositional image caption evaluation.
\newblock In \emph{Proceedings of the European Conference on Computer Vision (ECCV)}, pages 382--398, 2016.

\bibitem[Bai et~al.(2025)Bai, Chen, Liu, Wang, Ge, Song, Dang, Wang, Wang, Tang, et~al.]{bai2025qwen2}
Shuai Bai, Keqin Chen, Xuejing Liu, Jialin Wang, Wenbin Ge, Sibo Song, Kai Dang, Peng Wang, Shijie Wang, Jun Tang, et~al.
\newblock {Qwen2.5-VL Technical Report}.
\newblock \emph{arXiv preprint arXiv:2502.13923}, 2025.

\bibitem[Banerjee and Lavie(2005)]{banerjee2005meteor}
Satanjeev Banerjee and Alon Lavie.
\newblock {METEOR}: An automatic metric for {MT} evaluation with improved correlation with human judgments.
\newblock In \emph{Proceedings of the {ACL} Workshop on Intrinsic and Extrinsic Evaluation Measures for Machine Translation and/or Summarization}, pages 65--72, 2005.

\bibitem[Bińkowski et~al.(2018)Bińkowski, Sutherland, Arbel, and Gretton]{binkowski2018demystifying}
Mikołaj Bińkowski, Dougal~J. Sutherland, Michael Arbel, and Arthur Gretton.
\newblock Demystifying {MMD} {GAN}s.
\newblock In \emph{Proceedings of the International Conference on Learning Representations (ICLR)}, 2018.

\bibitem[Bukowski and S{\'e}quin(1995)]{bukowski1995object}
Richard~W Bukowski and Carlo~H. S{\'e}quin.
\newblock Object associations: A simple and practical approach to virtual {3D} manipulation.
\newblock In \emph{Proceedings of the 1995 Symposium on Interactive 3D Graphics}, pages 131--ff., 1995.

\bibitem[{\c{C}}elen et~al.(2024){\c{C}}elen, Han, Schindler, Van~Gool, Armeni, Obukhov, and Wang]{ccelen2024design}
Ata {\c{C}}elen, Guo Han, Konrad Schindler, Luc Van~Gool, Iro Armeni, Anton Obukhov, and Xi Wang.
\newblock {I-Design}: Personalized {LLM} interior designer.
\newblock \emph{arXiv preprint arXiv:2404.02838}, 2024.

\bibitem[Chang et~al.(2015)Chang, Monroe, Savva, Potts, and Manning]{chang2015text}
Angel Chang, Will Monroe, Manolis Savva, Christopher Potts, and Christopher~D Manning.
\newblock Text to {3D} scene generation with rich lexical grounding.
\newblock In \emph{Proceedings of the 53rd Annual Meeting of the Association for Computational Linguistics and the 7th International Joint Conference on Natural Language Processing (Volume 1: Long Papers)}, pages 53--62, 2015.

\bibitem[Clay and Wilhelms(1996)]{clay1996put}
Sharon~Rose Clay and Jane Wilhelms.
\newblock Put: Language-based interactive manipulation of objects.
\newblock \emph{IEEE Computer Graphics and Applications}, 16\penalty0 (2):\penalty0 31--39, 1996.

\bibitem[Coyne and Sproat(2001)]{coyne2001wordseye}
Bob Coyne and Richard Sproat.
\newblock {WordsEye}: An automatic text-to-scene conversion system.
\newblock In \emph{Proceedings of the 28th Annual Conference on Computer Graphics and Interactive Techniques}, pages 487--496, 2001.

\bibitem[Deitke et~al.(2022)Deitke, VanderBilt, Herrasti, Weihs, Ehsani, Salvador, Han, Kolve, Kembhavi, and Mottaghi]{deitke2022procthor}
Matt Deitke, Eli VanderBilt, Alvaro Herrasti, Luca Weihs, Kiana Ehsani, Jordi Salvador, Winson Han, Eric Kolve, Aniruddha Kembhavi, and Roozbeh Mottaghi.
\newblock {ProcTHOR}: Large-scale embodied {AI} using procedural generation.
\newblock In \emph{Advances in Neural Information Processing Systems}, pages 5982--5994, 2022.

\bibitem[Deitke et~al.(2023)Deitke, Schwenk, Salvador, Weihs, Michel, VanderBilt, Schmidt, Ehsani, Kembhavi, and Farhadi]{deitke2023objaverse}
Matt Deitke, Dustin Schwenk, Jordi Salvador, Luca Weihs, Oscar Michel, Eli VanderBilt, Ludwig Schmidt, Kiana Ehsani, Aniruddha Kembhavi, and Ali Farhadi.
\newblock Objaverse: A universe of annotated {3D} objects.
\newblock In \emph{Proceedings of the IEEE Conference on Computer Vision and Pattern Recognition (CVPR)}, pages 13142--13153, 2023.

\bibitem[Duan et~al.(2025)Duan, Yu, Chen, Fei-Fei, and Wu]{duan2025worldscore}
Haoyi Duan, Hong-Xing Yu, Sirui Chen, Li Fei-Fei, and Jiajun Wu.
\newblock {WorldScore: A unified evaluation benchmark for world generation}.
\newblock \emph{arXiv preprint arXiv:2504.00983}, 2025.

\bibitem[Feng et~al.(2023)Feng, Zhu, Fu, Jampani, Akula, He, Basu, Wang, and Wang]{feng2024layoutgpt}
Weixi Feng, Wanrong Zhu, Tsu-jui Fu, Varun Jampani, Arjun Akula, Xuehai He, Sugato Basu, Xin~Eric Wang, and William~Yang Wang.
\newblock {LayoutGPT}: Compositional visual planning and generation with large language models.
\newblock In \emph{Advances in Neural Information Processing Systems}, pages 18225--18250, 2023.

\bibitem[Fisher et~al.(2012)Fisher, Ritchie, Savva, Funkhouser, and Hanrahan]{fisher2012example}
Matthew Fisher, Daniel Ritchie, Manolis Savva, Thomas Funkhouser, and Pat Hanrahan.
\newblock Example-based synthesis of {3D} object arrangements.
\newblock \emph{ACM Transactions on Graphics (TOG)}, 31\penalty0 (6):\penalty0 1--11, 2012.

\bibitem[Fu et~al.(2021{\natexlab{a}})Fu, Cai, Gao, Zhang, Wang, Li, Zeng, Sun, Jia, Zhao, et~al.]{fu20213dfront}
Huan Fu, Bowen Cai, Lin Gao, Ling-Xiao Zhang, Jiaming Wang, Cao Li, Qixun Zeng, Chengyue Sun, Rongfei Jia, Binqiang Zhao, et~al.
\newblock {3D-FRONT}: {3D} furnished rooms with layouts and semantics.
\newblock In \emph{Proceedings of the IEEE International Conference on Computer Vision (ICCV)}, pages 10913--10922, 2021{\natexlab{a}}.

\bibitem[Fu et~al.(2021{\natexlab{b}})Fu, Jia, Gao, Gong, Zhao, Maybank, and Tao]{fu20213dfuture}
Huan Fu, Rongfei Jia, Lin Gao, Mingming Gong, Binqiang Zhao, Steve Maybank, and Dacheng Tao.
\newblock {3D-FUTURE}: {3D} furniture shape with texture.
\newblock \emph{International Journal of Computer Vision (IJCV)}, 129:\penalty0 3313--3337, 2021{\natexlab{b}}.

\bibitem[Fu et~al.(2024)Fu, Wen, Liu, and Sridhar]{fu2024anyhome}
Rao Fu, Zehao Wen, Zichen Liu, and Srinath Sridhar.
\newblock {AnyHome}: Open-vocabulary generation of structured and textured {3D} homes.
\newblock In \emph{Proceedings of the European Conference on Computer Vision (ECCV)}, pages 52--70, 2024.

\bibitem[Gu et~al.(2025{\natexlab{a}})Gu, Huang, Je, Yang, and Guibas]{gu2025blendergym}
Yunqi Gu, Ian Huang, Jihyeon Je, Guandao Yang, and Leonidas Guibas.
\newblock {BlenderGym: Benchmarking Foundational Model Systems for Graphics Editing}.
\newblock \emph{arXiv preprint arXiv:2504.01786}, 2025{\natexlab{a}}.

\bibitem[Gu et~al.(2025{\natexlab{b}})Gu, Cui, Li, Wei, Ge, Gu, Liu, Davis, and Ding]{gu2025artiscene}
Zeqi Gu, Yin Cui, Zhaoshuo Li, Fangyin Wei, Yunhao Ge, Jinwei Gu, Ming-Yu Liu, Abe Davis, and Yifan Ding.
\newblock {ArtiScene}: Language-driven artistic {3D} scene generation through image intermediary.
\newblock In \emph{Proceedings of the Computer Vision and Pattern Recognition Conference}, pages 2891--2901, 2025{\natexlab{b}}.

\bibitem[Hessel et~al.(2021)Hessel, Holtzman, Forbes, Bras, and Choi]{hessel2021clipscore}
Jack Hessel, Ari Holtzman, Maxwell Forbes, Ronan~Le Bras, and Yejin Choi.
\newblock {CLIPScore}: A reference-free evaluation metric for image captioning.
\newblock In \emph{Proceedings of the Conference on Empirical Methods in Natural Language Processing (EMNLP)}, 2021.

\bibitem[Heusel et~al.(2017)Heusel, Ramsauer, Unterthiner, Nessler, and Hochreiter]{heusel2017gans}
Martin Heusel, Hubert Ramsauer, Thomas Unterthiner, Bernhard Nessler, and Sepp Hochreiter.
\newblock {GANs} trained by a two time-scale update rule converge to a local nash equilibrium.
\newblock In \emph{Advances in Neural Information Processing Systems}, 2017.

\bibitem[Hu et~al.(2024{\natexlab{a}})Hu, Arroyo, Debats, Manhardt, Carlone, and Tombari]{hu2024mixed}
Siyi Hu, Diego~Martin Arroyo, Stephanie Debats, Fabian Manhardt, Luca Carlone, and Federico Tombari.
\newblock Mixed diffusion for {3D} indoor scene synthesis.
\newblock \emph{arXiv preprint arXiv:2405.21066}, 2024{\natexlab{a}}.

\bibitem[Hu et~al.(2024{\natexlab{b}})Hu, Iscen, Jain, Kipf, Yue, Ross, Schmid, and Fathi]{hu2024scenecraft}
Ziniu Hu, Ahmet Iscen, Aashi Jain, Thomas Kipf, Yisong Yue, David~A Ross, Cordelia Schmid, and Alireza Fathi.
\newblock {SceneCraft}: An {LLM} agent for synthesizing {3D} scenes as {Blender} code.
\newblock In \emph{Proceedings of the International Conference on Machine Learning (ICML)}, pages 19252--19282, 2024{\natexlab{b}}.

\bibitem[Huang et~al.(2025)Huang, Zhai, Bauer, Pollefeys, Tombari, Guibas, Huang, and Engelmann]{huang2025video}
Rui Huang, Guangyao Zhai, Zuria Bauer, Marc Pollefeys, Federico Tombari, Leonidas Guibas, Gao Huang, and Francis Engelmann.
\newblock Video perception models for {3D} scene synthesis.
\newblock \emph{arXiv preprint arXiv:2506.20601}, 2025.

\bibitem[Huang et~al.(2024{\natexlab{a}})Huang, He, Yu, Zhang, Si, Jiang, Zhang, Wu, Jin, Chanpaisit, et~al.]{huang2024vbench}
Ziqi Huang, Yinan He, Jiashuo Yu, Fan Zhang, Chenyang Si, Yuming Jiang, Yuanhan Zhang, Tianxing Wu, Qingyang Jin, Nattapol Chanpaisit, et~al.
\newblock {VBench}: Comprehensive benchmark suite for video generative models.
\newblock In \emph{Proceedings of the IEEE Conference on Computer Vision and Pattern Recognition (CVPR)}, pages 21807--21818, 2024{\natexlab{a}}.

\bibitem[Huang et~al.(2024{\natexlab{b}})Huang, Zhang, Xu, He, Yu, Dong, Ma, Chanpaisit, Si, Jiang, et~al.]{huang2024vbench++}
Ziqi Huang, Fan Zhang, Xiaojie Xu, Yinan He, Jiashuo Yu, Ziyue Dong, Qianli Ma, Nattapol Chanpaisit, Chenyang Si, Yuming Jiang, et~al.
\newblock {VBench++}: Comprehensive and versatile benchmark suite for video generative models.
\newblock \emph{arXiv preprint arXiv:2411.13503}, 2024{\natexlab{b}}.

\bibitem[Keshavarzi et~al.(2020)Keshavarzi, Parikh, Zhai, Mao, Caldas, and Yang]{keshavarzi2020scenegen}
Mohammad Keshavarzi, Aakash Parikh, Xiyu Zhai, Melody Mao, Luisa Caldas, and Allen~Y Yang.
\newblock {SceneGen}: Generative contextual scene augmentation using scene graph priors.
\newblock \emph{arXiv preprint arXiv:2009.12395}, 2020.

\bibitem[Krishna et~al.(2017)Krishna, Zhu, Groth, Johnson, Hata, Kravitz, Chen, Kalantidis, Li, Shamma, et~al.]{krishna2017visual}
Ranjay Krishna, Yuke Zhu, Oliver Groth, Justin Johnson, Kenji Hata, Joshua Kravitz, Stephanie Chen, Yannis Kalantidis, Li-Jia Li, David~A Shamma, et~al.
\newblock Visual genome: Connecting language and vision using crowdsourced dense image annotations.
\newblock \emph{International Journal of Computer Vision (IJCV)}, 123\penalty0 (1):\penalty0 32--73, 2017.

\bibitem[Kynk{\"a}{\"a}nniemi et~al.(2023)Kynk{\"a}{\"a}nniemi, Karras, Aittala, Aila, and Lehtinen]{kynkaanniemi2022role}
Tuomas Kynk{\"a}{\"a}nniemi, Tero Karras, Miika Aittala, Timo Aila, and Jaakko Lehtinen.
\newblock The role of {ImageNet} classes in {F}r{\'{e}}chet inception distance.
\newblock In \emph{Proceedings of the International Conference on Learning Representations (ICLR)}, 2023.

\bibitem[Li et~al.(2023)Li, Li, Savarese, and Hoi]{li2023blip}
Junnan Li, Dongxu Li, Silvio Savarese, and Steven Hoi.
\newblock {BLIP-2}: Bootstrapping language-image pre-training with frozen image encoders and large language models.
\newblock In \emph{International conference on machine learning}, pages 19730--19742. PMLR, 2023.

\bibitem[Li et~al.(2019)Li, Patil, Xu, Chaudhuri, Khan, Shamir, Tu, Chen, Cohen-Or, and Zhang]{li2019grains}
Manyi Li, Akshay~Gadi Patil, Kai Xu, Siddhartha Chaudhuri, Owais Khan, Ariel Shamir, Changhe Tu, Baoquan Chen, Daniel Cohen-Or, and Hao Zhang.
\newblock {GRAINS}: Generative recursive autoencoders for indoor scenes.
\newblock \emph{ACM Transactions on Graphics (TOG)}, 38\penalty0 (2):\penalty0 1--16, 2019.

\bibitem[Lin and MU(2024)]{lin2024instructscene}
Chenguo Lin and Yadong MU.
\newblock {InstructScene}: Instruction-driven {3D} indoor scene synthesis with semantic graph prior.
\newblock In \emph{Proceedings of the International Conference on Learning Representations (ICLR)}, 2024.

\bibitem[Lin(2004)]{lin2004rouge}
Chin-Yew Lin.
\newblock {ROUGE}: A package for automatic evaluation of summaries.
\newblock In \emph{Text Summarization Branches Out}, pages 74--81, 2004.

\bibitem[Lin et~al.(2024)Lin, Pathak, Li, Li, Xia, Neubig, Zhang, and Ramanan]{lin2025evaluating}
Zhiqiu Lin, Deepak Pathak, Baiqi Li, Jiayao Li, Xide Xia, Graham Neubig, Pengchuan Zhang, and Deva Ramanan.
\newblock Evaluating text-to-visual generation with image-to-text generation.
\newblock In \emph{Proceedings of the European Conference on Computer Vision (ECCV)}, pages 366--384, 2024.

\bibitem[Ling et~al.(2025)Ling, Lin, Lin, Ding, Zeng, Sheng, Ge, Liu, Bera, and Li]{ling2025scenethesis}
Lu Ling, Chen-Hsuan Lin, Tsung-Yi Lin, Yifan Ding, Yu Zeng, Yichen Sheng, Yunhao Ge, Ming-Yu Liu, Aniket Bera, and Zhaoshuo Li.
\newblock Scenethesis: A language and vision agentic framework for {3D} scene generation.
\newblock \emph{arXiv preprint arXiv:2505.02836}, 2025.

\bibitem[Ma et~al.(2018)Ma, Patil, Fisher, Li, Pirk, Hua, Yeung, Tong, Guibas, and Zhang]{ma2018language}
Rui Ma, Akshay~Gadi Patil, Matthew Fisher, Manyi Li, S{\"o}ren Pirk, Binh-Son Hua, Sai-Kit Yeung, Xin Tong, Leonidas Guibas, and Hao Zhang.
\newblock Language-driven synthesis of {3D} scenes from scene databases.
\newblock \emph{ACM Transactions on Graphics (TOG)}, 37\penalty0 (6):\penalty0 1--16, 2018.

\bibitem[OpenAI(2025)]{openai2025gpto4}
OpenAI.
\newblock {OpenAI o3 and o4-mini System Card}.
\newblock 2025.

\bibitem[Papineni et~al.(2002)Papineni, Roukos, Ward, and Zhu]{papineni2002bleu}
Kishore Papineni, Salim Roukos, Todd Ward, and Wei-Jing Zhu.
\newblock {BLEU}: A method for automatic evaluation of machine translation.
\newblock In \emph{Proceedings of the 40th annual meeting of the Association for Computational Linguistics}, pages 311--318, 2002.

\bibitem[Paschalidou et~al.(2021)Paschalidou, Kar, Shugrina, Kreis, Geiger, and Fidler]{paschalidou2021atiss}
Despoina Paschalidou, Amlan Kar, Maria Shugrina, Karsten Kreis, Andreas Geiger, and Sanja Fidler.
\newblock {ATISS}: Autoregressive transformers for indoor scene synthesis.
\newblock In \emph{Advances in Neural Information Processing Systems}, pages 12013--12026, 2021.

\bibitem[Pfaff et~al.(2025)Pfaff, Dai, Zakharov, Iwase, and Tedrake]{pfaff2025steerable}
Nicholas Pfaff, Hongkai Dai, Sergey Zakharov, Shun Iwase, and Russ Tedrake.
\newblock Steerable scene generation with post training and inference-time search.
\newblock \emph{arXiv preprint arXiv:2505.04831}, 2025.

\bibitem[Raistrick et~al.(2024)Raistrick, Mei, Kayan, Yan, Zuo, Han, Wen, Parakh, Alexandropoulos, Lipson, et~al.]{raistrick2024infinigen}
Alexander Raistrick, Lingjie Mei, Karhan Kayan, David Yan, Yiming Zuo, Beining Han, Hongyu Wen, Meenal Parakh, Stamatis Alexandropoulos, Lahav Lipson, et~al.
\newblock {Infinigen Indoors}: Photorealisltic indoor scenes using procedural generation.
\newblock In \emph{Proceedings of the IEEE Conference on Computer Vision and Pattern Recognition (CVPR)}, pages 21783--21794, 2024.

\bibitem[Ritchie et~al.(2019)Ritchie, Wang, and Lin]{ritchie2019fast}
Daniel Ritchie, Kai Wang, and Yu-an Lin.
\newblock Fast and flexible indoor scene synthesis via deep convolutional generative models.
\newblock In \emph{Proceedings of the IEEE Conference on Computer Vision and Pattern Recognition (CVPR)}, pages 6175--6183, 2019.

\bibitem[Savva et~al.(2016)Savva, Chang, Hanrahan, Fisher, and Nie{\ss}ner]{savva2016pigraphs}
Manolis Savva, Angel~X Chang, Pat Hanrahan, Matthew Fisher, and Matthias Nie{\ss}ner.
\newblock Pigraphs: Learning interaction snapshots from observations.
\newblock \emph{ACM Transactions on Graphics (TOG)}, 35\penalty0 (4):\penalty0 1--12, 2016.

\bibitem[Schuster et~al.(2015)Schuster, Krishna, Chang, Fei-Fei, and Manning]{schuster2015generating}
Sebastian Schuster, Ranjay Krishna, Angel Chang, Li Fei-Fei, and Christopher~D Manning.
\newblock Generating semantically precise scene graphs from textual descriptions for improved image retrieval.
\newblock In \emph{Proceedings of the fourth workshop on vision and language}, pages 70--80, 2015.

\bibitem[Shinya and Forgue(1995)]{shinya1995laying}
Mikio Shinya and Marie-Claire Forgue.
\newblock Laying out objects with geometric and physical constraints.
\newblock \emph{The Visual Computer}, 11:\penalty0 188--201, 1995.

\bibitem[Sun et~al.(2025{\natexlab{a}})Sun, Liu, Gu, Lim, Bhat, Tombari, Li, Haber, and Wu]{sun2025layoutvlm}
Fan-Yun Sun, Weiyu Liu, Siyi Gu, Dylan Lim, Goutam Bhat, Federico Tombari, Manling Li, Nick Haber, and Jiajun Wu.
\newblock {LayoutVLM}: Differentiable optimization of {3D} layout via vision-language models.
\newblock In \emph{Proceedings of the IEEE Conference on Computer Vision and Pattern Recognition (CVPR)}, pages 29469--29478, 2025{\natexlab{a}}.

\bibitem[Sun et~al.(2024)Sun, Zhou, Zhou, Li, and Li]{sun2025forest2seq}
Qi Sun, Hang Zhou, Wengang Zhou, Li Li, and Houqiang Li.
\newblock {Forest2Seq}: Revitalizing order prior for sequential indoor scene synthesis.
\newblock In \emph{Proceedings of the European Conference on Computer Vision (ECCV)}, pages 251--268, 2024.

\bibitem[Sun et~al.(2025{\natexlab{b}})Sun, Goel, and Chang]{sun2025semlayoutdiff}
Xiaohao Sun, Divyam Goel, and Angel~X Chang.
\newblock Semlayoutdiff: Semantic layout generation with diffusion model for indoor scene synthesis.
\newblock \emph{arXiv preprint arXiv:2508.18597}, 2025{\natexlab{b}}.

\bibitem[Tang et~al.(2024)Tang, Nie, Markhasin, Dai, Thies, and Nie{\ss}ner]{tang2024diffuscene}
Jiapeng Tang, Yinyu Nie, Lev Markhasin, Angela Dai, Justus Thies, and Matthias Nie{\ss}ner.
\newblock {DiffuScene}: Denoising diffusion models for generative indoor scene synthesis.
\newblock In \emph{Proceedings of the IEEE Conference on Computer Vision and Pattern Recognition (CVPR)}, pages 20507--20518, 2024.

\bibitem[Vedantam et~al.(2015)Vedantam, Lawrence~Zitnick, and Parikh]{vedantam2015cider}
Ramakrishna Vedantam, C Lawrence~Zitnick, and Devi Parikh.
\newblock {CIDEr}: Consensus-based image description evaluation.
\newblock In \emph{Proceedings of the IEEE Conference on Computer Vision and Pattern Recognition (CVPR)}, pages 4566--4575, 2015.

\bibitem[Wang et~al.(2024{\natexlab{a}})Wang, Zhong, Chai, He, Chen, and Liao]{wang2024chat2layout}
Can Wang, Hongliang Zhong, Menglei Chai, Mingming He, Dongdong Chen, and Jing Liao.
\newblock {Chat2Layout}: Interactive {3D} furniture layout with a multimodal {LLM}.
\newblock \emph{arXiv preprint arXiv:2407.21333}, 2024{\natexlab{a}}.

\bibitem[Wang et~al.(2018)Wang, Savva, Chang, and Ritchie]{wang2018deep}
Kai Wang, Manolis Savva, Angel~X Chang, and Daniel Ritchie.
\newblock Deep convolutional priors for indoor scene synthesis.
\newblock \emph{ACM Transactions on Graphics (TOG)}, 37\penalty0 (4):\penalty0 1--14, 2018.

\bibitem[Wang et~al.(2019)Wang, Lin, Weissmann, Savva, Chang, and Ritchie]{wang2019planit}
Kai Wang, Yu-An Lin, Ben Weissmann, Manolis Savva, Angel~X Chang, and Daniel Ritchie.
\newblock {PlanIT}: Planning and instantiating indoor scenes with relation graph and spatial prior networks.
\newblock \emph{ACM Transactions on Graphics (TOG)}, 38\penalty0 (4):\penalty0 1--15, 2019.

\bibitem[Wang et~al.(2024{\natexlab{b}})Wang, Qiu, Liu, Chen, Cai, Wang, Wang, Xian, and Gan]{wang2024architect}
Yian Wang, Xiaowen Qiu, Jiageng Liu, Zhehuan Chen, Jiting Cai, Yufei Wang, Tsun-Hsuan Wang, Zhou Xian, and Chuang Gan.
\newblock Architect: Generating vivid and interactive {3D} scenes with hierarchical {2D} inpainting.
\newblock In \emph{Advances in Neural Information Processing Systems}, pages 67575--67603, 2024{\natexlab{b}}.

\bibitem[Wu et~al.(2024)Wu, Yang, Li, Zhang, Liu, Guibas, Lin, and Wetzstein]{wu2024gpt}
Tong Wu, Guandao Yang, Zhibing Li, Kai Zhang, Ziwei Liu, Leonidas Guibas, Dahua Lin, and Gordon Wetzstein.
\newblock {GPT-4V}(ision) is a human-aligned evaluator for text-to-{3D} generation.
\newblock In \emph{Proceedings of the IEEE Conference on Computer Vision and Pattern Recognition (CVPR)}, pages 22227--22238, 2024.

\bibitem[Yang et~al.(2024{\natexlab{a}})Yang, Jia, Zhi, and Huang]{yang2024physcene}
Yandan Yang, Baoxiong Jia, Peiyuan Zhi, and Siyuan Huang.
\newblock {PhyScene}: Physically interactable {3D} scene synthesis for embodied {AI}.
\newblock In \emph{Proceedings of the IEEE Conference on Computer Vision and Pattern Recognition (CVPR)}, pages 16262--16272, 2024{\natexlab{a}}.

\bibitem[Yang et~al.(2024{\natexlab{b}})Yang, Sun, Weihs, VanderBilt, Herrasti, Han, Wu, Haber, Krishna, Liu, et~al.]{yang2024holodeck}
Yue Yang, Fan-Yun Sun, Luca Weihs, Eli VanderBilt, Alvaro Herrasti, Winson Han, Jiajun Wu, Nick Haber, Ranjay Krishna, Lingjie Liu, et~al.
\newblock Holodeck: Language guided generation of {3D} embodied {AI} environments.
\newblock In \emph{Proceedings of the IEEE Conference on Computer Vision and Pattern Recognition (CVPR)}, pages 16277--16287, 2024{\natexlab{b}}.

\bibitem[Ye et~al.(2024)Ye, Zheng, Liu, and Peng]{ye2024relscene}
Zhaoda Ye, Xinhan Zheng, Yang Liu, and Yuxin Peng.
\newblock {RelScene}: A benchmark and baseline for spatial relations in text-driven {3D} scene generation.
\newblock In \emph{Proceedings of the 32nd ACM International Conference on Multimedia}, pages 10563--10571, 2024.

\bibitem[Yeh et~al.(2012)Yeh, Yang, Watson, Goodman, and Hanrahan]{yeh2012synthesizing}
Yi-Ting Yeh, Lingfeng Yang, Matthew Watson, Noah~D Goodman, and Pat Hanrahan.
\newblock Synthesizing open worlds with constraints using locally annealed reversible jump {MCMC}.
\newblock \emph{ACM Transactions on Graphics (TOG)}, 31\penalty0 (4):\penalty0 1--11, 2012.

\bibitem[Yu et~al.(2011)Yu, Yeung, Tang, Terzopoulos, Chan, and Osher]{yu2011make}
Lap~Fai Yu, Sai~Kit Yeung, Chi~Keung Tang, Demetri Terzopoulos, Tony~F Chan, and Stanley~J Osher.
\newblock {Make it Home}: Automatic optimization of furniture arrangement.
\newblock \emph{ACM Transactions on Graphics (TOG)}, 30\penalty0 (4), 2011.

\bibitem[Zhai et~al.(2023)Zhai, {\"O}rnek, Wu, Di, Tombari, Navab, and Busam]{zhai2024commonscenes}
Guangyao Zhai, Evin~P{\i}nar {\"O}rnek, Shun-Cheng Wu, Yan Di, Federico Tombari, Nassir Navab, and Benjamin Busam.
\newblock {CommonScenes}: Generating commonsense {3D} indoor scenes with scene graphs.
\newblock In \emph{Advances in Neural Information Processing Systems}, pages 30026--30038, 2023.

\bibitem[Zhai et~al.(2024)Zhai, {\"O}rnek, Chen, Liao, Di, Navab, Tombari, and Busam]{zhai2024echoscene}
Guangyao Zhai, Evin~P{\i}nar {\"O}rnek, Dave~Zhenyu Chen, Ruotong Liao, Yan Di, Nassir Navab, Federico Tombari, and Benjamin Busam.
\newblock {EchoScene}: Indoor scene generation via information echo over scene graph diffusion.
\newblock In \emph{Proceedings of the European Conference on Computer Vision (ECCV)}, pages 167--184, 2024.

\bibitem[Zhang et~al.(2024)Zhang, Zhang, Dong, Zang, and Wang]{zhang2024longclip}
Beichen Zhang, Pan Zhang, Xiaoyi Dong, Yuhang Zang, and Jiaqi Wang.
\newblock {Long-CLIP}: Unlocking the long-text capability of {CLIP}.
\newblock In \emph{Proceedings of the European Conference on Computer Vision (ECCV)}, page 310–325, 2024.

\bibitem[Zhang et~al.(2020)Zhang, Kishore, Wu, Weinberger, and Artzi]{zhang2019bertscore}
Tianyi Zhang, Varsha Kishore, Felix Wu, Kilian~Q Weinberger, and Yoav Artzi.
\newblock {BERTScore}: Evaluating text generation with {BERT}.
\newblock In \emph{Proceedings of the International Conference on Learning Representations (ICLR)}, 2020.

\bibitem[Zhang et~al.(2025)Zhang, Zhang, Wu, Wang, Wetzstein, Lin, and Liu]{zhang20253dgen}
Yuhan Zhang, Mengchen Zhang, Tong Wu, Tengfei Wang, Gordon Wetzstein, Dahua Lin, and Ziwei Liu.
\newblock {3DGen-Bench: Comprehensive Benchmark Suite for 3D Generative Models}.
\newblock \emph{arXiv preprint arXiv:2503.21745}, 2025.

\end{thebibliography}
}

\maketitlesupplementary
\appendix
\section*{Appendices}
In this appendix, we provide details about our \ourdataset dataset in \cref{sec:supp_dataset_details},
the implementation details of our metrics in \cref{sec:supp_metric_implementation_details},
experimental results of running \ours with an open-source VLM in \cref{sec:supp_qwen},
and limitations of our work in \cref{sec:supp_limitations}.
We also provide the details of our user study in \cref{sec:supp_user_study},
the vision language model (VLM) prompts we used in \cref{sec:supp_llm_prompts},
and information about scientific artifacts involved and AI assistant usage in this work in \cref{sec:supp_scientific_artifacts,sec:supp_ai_assistants_usage}.

\begin{table*}
\centering
\resizebox{\linewidth}{!}
{
\begin{tabular}{@{} ll rrrr rrrrrr r@{}}
\toprule
&& \multicolumn{4}{c}{Fidelity} & \multicolumn{6}{c}{Plausbility}
\\ 
  
\cmidrule(l){3-6} \cmidrule(l){7-12}
                                        & Difficulty
                                        & $\uparrow~$CNT$_{\%}$ & $\uparrow~$ATR$_{\%}$ & $\uparrow~$OOR$_{\%}$ & $\uparrow~$OAR$_{\%}$
                                        & $\downarrow~$COL$_{ob\%}$ & $\downarrow~$COL$_{sc\%}$ & $\uparrow~$SUP$_{\%}$
                                        & $\uparrow~$NAV$_{\%}$ & $\uparrow~$ACC$_{\%}$ & $\downarrow~$OOB$_{\%}$
\\
\midrule

\multirow{3}{*}{ATISS} &
Easy &
18.72 & 13.00 & 2.49 & 8.40 &
53.12 & 79.33 & 90.38 &
99.92 & 84.56 & 11.38
\\ &
Medium &
12.93 & 8.32 & 1.46 & 7.81 &
50.53 & 73.50 & 91.50 &
99.80 & 85.81 & 10.14
\\ &
Hard &
8.09 & 5.13 & 0.65 & 8.00 &
47.27 & 73.33 & 90.65 &
99.76 & 86.22 & 11.30
\\ \midrule

\multirow{3}{*}{DiffuScene} &
Easy &
20.64 & 15.20 & 4.48 & \textsuperscript{\textdagger}8.00 &
33.52 & 64.67 & 74.48 &
\textsuperscript{\textdagger}99.51 & 81.32 & \textsuperscript{\textdagger}26.56
\\ &
Medium &
14.26 & 11.03 & 4.26 & \textsuperscript{\textdagger}9.31 &
31.42 & 62.50 & 75.10 &
\textsuperscript{\textdagger}99.41 & 82.72 & \textsuperscript{\textdagger}25.29
\\ &
Hard &
8.25 & 6.45 & 2.42 & \textsuperscript{\textdagger}7.62 &
30.50 & 61.33 & 76.80 &
\textsuperscript{\textdagger}99.41 & 81.32 & \textsuperscript{\textdagger}25.63
\\ \midrule

\multirow{3}{*}{LayoutGPT} &
Easy &
24.47 & 15.75 & 3.98 & 4.40 &
12.50 & 30.00 & 31.85 &
99.98 & 48.34 & 71.75
\\ &
Medium &
14.18 & 9.37 & 1.46 & 5.11 &
11.07 & 25.50 & 29.45 &
100.00 & 48.18 & 72.38
\\ &
Hard &
7.04 & 4.90 & 0.65 & 4.95 &
10.88 & 26.67 & 29.19 &
100.00 & 44.86 & 72.60
\\ \midrule

\multirow{3}{*}{InstructScene} &
Easy &
23.19 & 16.48 & 9.45 & \textsuperscript{\textdagger}10.40 &
56.11 & 89.33 & 77.58 &
\textsuperscript{\textdagger}98.41 & 77.14 & \textsuperscript{\textdagger}24.31

\\ &
Medium &
17.51 & 13.74 & 4.38 & \textsuperscript{\textdagger}12.01 &
52.64 & 84.00 & 80.55 &
\textsuperscript{\textdagger}98.86 & 78.02 & \textsuperscript{\textdagger}19.03
\\ &
Hard &
9.57 & 8.76 & 2.35 & \textsuperscript{\textdagger}8.95 &
56.92 & 84.67 & 84.69 &
\textsuperscript{\textdagger}99.74 & 77.11 & \textsuperscript{\textdagger}14.50
\\ \midrule

\multirow{3}{*}{LayoutVLM} &
Easy &
39.36 & 20.88 & 9.45 & 24.40 &
33.17 & 66.00 & 79.23 &
99.37 & 82.49 & 4.52
\\ &
Medium &
41.70 & 25.29 & 7.18 & 23.65 &
33.33 & 57.00 & 73.31 &
99.60 & 86.15 & 4.96
\\ &
Hard &
30.58 & 17.17 & 4.90 & 14.29 &
29.45 & 50.67 & 78.62 &
99.99 & 87.88 & 5.23
\\ \midrule

\multirow{3}{*}{Holodeck} &
Easy &
44.47 & 39.19 & 22.89 & 38.00 &
16.20 & 70.67 & 60.83 &
99.61 & 89.75 & 1.57
\\ &
Medium &
36.61 & 32.99 & 14.81 & 39.64 &
16.26 & 70.00 & 64.40 &
99.57 & 90.06 & 1.31
\\ &
Hard &
26.95 & 22.64 & 8.10 & 35.43 &
15.23 & 76.67 & 63.96 &
99.64 & 89.07 & 1.49
\\

\bottomrule
\end{tabular}
}
\vspace{-0.5em}
\caption{
Breakdown of evaluation results by difficulty using \ours with the \ourdataset dataset on six recent scene generation methods.
We report the values for each metric averaged across all scenes in each difficulty level.
\textsuperscript{\textdagger} indicates numbers related to the floor plan shape but the method cannot be conditioned on it.
As the difficulty increases, the performance of the methods at generating scenes with the correct number of objects decreases.
Compared to the other methods, Holodeck has the best overall performance across all difficulty levels.
}
\label{tab:result_by_difficulty}
\end{table*}

\section{Dataset Details}
\label{sec:supp_dataset_details}

We provide additional details about our \ourdataset dataset below, including the annotation schema (\cref{sec:supp_annotation_schema}) and the data collection process (\cref{sec:supp_data_collection}).

\subsection{Annotation Schema}
\label{sec:supp_annotation_schema}

\ourdataset contains four annotation fields: object count, object attribute, object-object relationships, and object-architecture relationships.
\cref{tab:annotation_schema} shows the annotation schema for each field along with examples.
We describe the schema below.

\mypara{Quantifier and quantity} are used to specify the \textit{count} of an annotation entry.
For example, they are used to specify the number of objects in the scene for the object count field and the number of objects that have a specific attribute for the object attribute field.
The quantifier can be one of the following: \texttt{eq} (equal), \texttt{gt} (greater than), \texttt{lt} (less than), \texttt{ge} (greater than or equal), or \texttt{le} (less than or equal).
The quantity is an non-negative integer.

\begin{figure}
    \centering
    \includegraphics[width=\linewidth]{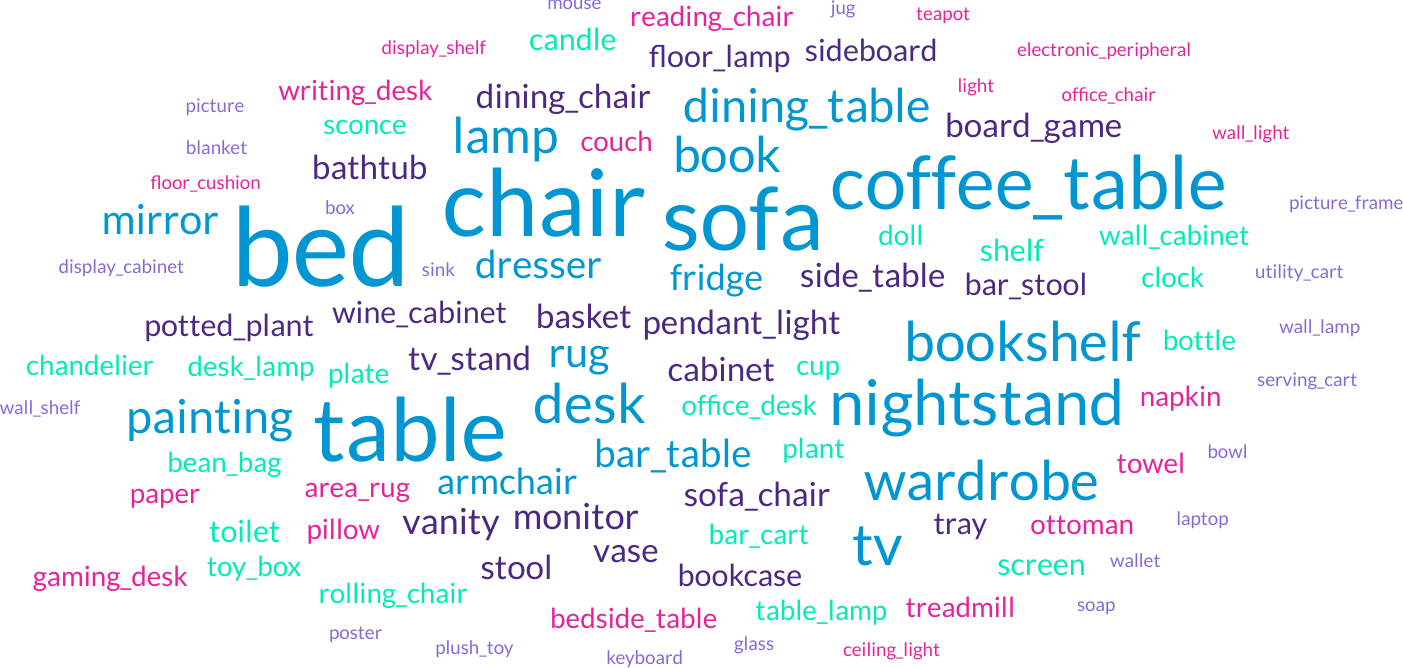}
    \vspace{-1.5em}
    \caption{
        Word cloud showing the most frequent object categories in \ourdataset.
    }
    \label{fig:category_word_cloud}
    \vspace{-1.5em}
\end{figure}

\mypara{Object category} specifies the object category of interest in an annotation entry.
It is an open-vocabulary string that specifies exactly one object category using a noun (e.g., bed) or a noun phrase (e.g., office chair).
In object-object relationships, multiple object categories are used to specify what objects are involved in a relationship, with the first object category being the anchor object that the relationship is based on.
\cref{fig:category_word_cloud} shows the most frequent object categories in \ourdataset.

\mypara{Attribute} specifies the attribute of an object in an object attribute annotation entry.
It is an open-vocabulary string that specifies exactly one attribute using an adjective.
For example, it can be color (e.g., red), material (e.g., wooden), shape (e.g., round), size (e.g., large), style (e.g., modern), or more specific attributes (e.g., queen-size).

\mypara{Architecture type} specifies the type of architectural element in an object-architecture relationship annotation entry.
It can be one of the following: \texttt{wall}, \texttt{floor}, \texttt{ceiling}, \texttt{window}, \texttt{door}, or \texttt{room}.
It can also be a more specific room type (e.g., bedroom, kitchen) for scenes with multiple rooms.

\mypara{Relationship} specifies the relationship between objects in an object-object relationship annotation entry and the relationship between an object and an architectural element in an object-architecture relationship annotation entry.
It is an open-vocabulary string that specifies exactly one relationship using a preposition (e.g., in front of, against), a verb (e.g., face, hang), or a prepositional phrase (e.g., at the foot of, at the corner of).

\subsection{Manual Data Collection Process}
\label{sec:supp_data_collection}

\cref{tab:example_descriptions} shows examples of scene descriptions and annotations in \ourdataset.
Below, we provide further details about manual curation of the initial 100 scene descriptions and annotations.

The first 100 scene descriptions and annotations in \ourdataset are written by the authors in English.
The annotators are primarily from Asia, aged between 20 and 30, have lived in Western countries, and speak English as their second language.
They have backgrounds in computer science, are familiar with the task of scene generation, and have been informed of the purpose of the dataset.

During the data collection process, the annotators were first given the definition of the difficulty levels, the annotation schema, and the target number of scenes to guide the annotation process.
They were then asked to write scene descriptions that are diverse and cover a wide range of object categories and relationships, drawing inspiration from their daily lives or from online sources.
At the same time, they were asked to annotate the scenes they wrote according to the annotation schema.
Both the scene descriptions and annotations were validated by the authors to ensure quality and consistency.
In addition, the scene descriptions were passed through an VLM to check for grammatical errors and typos.
No personal identifiable information was collected during the whole data collection process.

\begin{figure*}
    \centering
    \includegraphics[width=\linewidth]{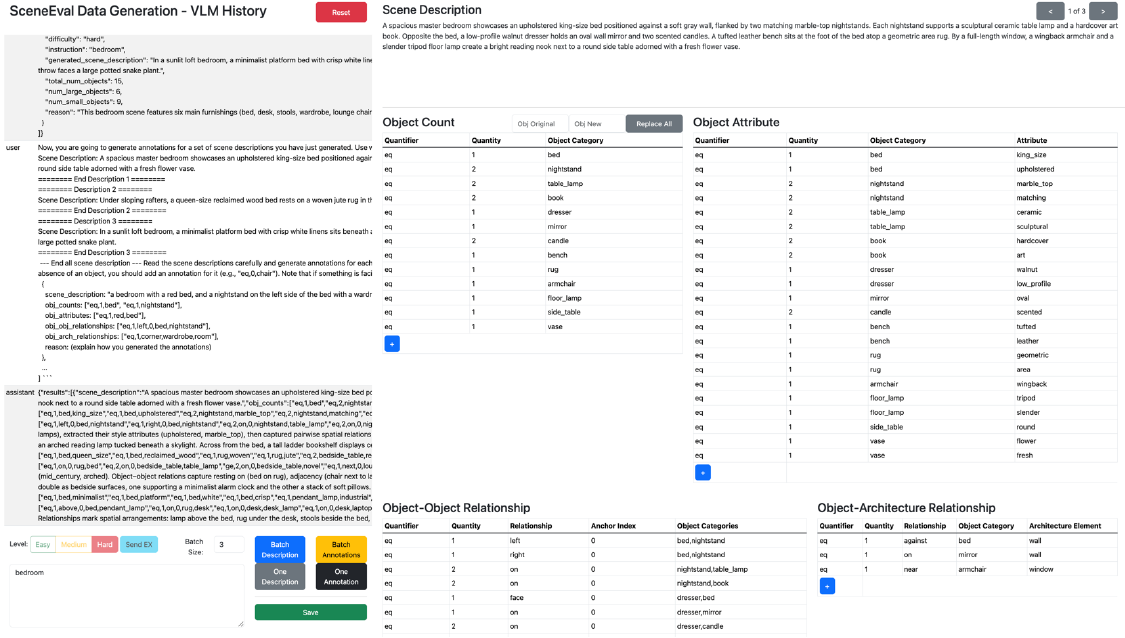}
    \vspace{-1.5em}
    \caption{
        Interface for semi-automatic data generation.
        \textbf{Top left:} Conversation history with the VLM.
        \textbf{Bottom left:} Controls for sending in-context examples and generating scene descriptions and annotations.
        \textbf{Top right:} Editable textbox containing the generated scene description.
        \textbf{Bottom right:} Editable tables containing the corresponding generated annotations.
        Each generated scene-description and annotation pair is manually validated before saving.
    }
    \label{fig:data_generation_ui}
    \vspace{-1em}
\end{figure*}

\begin{table}
    \centering
    \resizebox{\linewidth}{!}
    {
    \begin{tabular}{@{} ll @{}}
    \toprule
    Difficulty & Scene Description
    \\ \midrule
    Easy & \makecell[cl]{
        A simple bedroom featuring a twin bed against the wall, \\
        with a wardrobe positioned in the corner of the room, \\
        and a desk next to the window.
    } \\
    \midrule
    Medium & \makecell[cl]{
        This entertaining basement layout features a large gaming \\
        setup with two monitors on a desk against one wall, while \\
        a comfy bean bag chair is positioned nearby for casual \\
        seating. Across from the gaming area, a small cabinet \\
        with a mini fridge and a popcorn machine on top \\
        completes the setup.
    } \\
    \midrule
    Hard & \makecell[cl]{
        This cozy bedroom features a full-size bed against a wall \\
        with two nightstands on each side where the left one has a \\
        small clock. A small desk sits in the corner with a \\
        comfortable chair for studying or working. Opposite the \\
        bed, a spacious dresser provides additional storage. \\
        Adjacent to the bedroom, the living room has a recliner \\
        and an ottoman facing a low coffee table with a small \\
        vase and magazines. A large bookshelf in the corner holds \\
        books and board games, while an wide couch provides \\
        space for gatherings. Just off the living room, a small \\
        gaming room with a gaming console and two beanbag \\
        chairs offers a dedicated space for entertainment, \\
        complete with a small shelf for controllers and headsets.
    } \\

    \bottomrule
    \end{tabular}
    }
    \vspace{-0.5em}
    \caption{
        Example scene descriptions for the three difficulty levels in \ourdataset.
        The easy description specifies three large furniture objects.
        The medium one specifies three large and four small objects.
        The hard description specifies 14 large and 13 small objects.
    }
    \label{tab:example_descriptions}
    \vspace{-1em}
\end{table}

\section{Metric Implementation Details}
\label{sec:supp_metric_implementation_details}

We provide details about about the object renderings used in \ours in \cref{sec:supp_object_renderings}, additional implementation details for our metrics in \cref{sec:supp_implementation_details}, and details about the predefined spatial relationships used in our object-object relationship and object-architecture relationship metrics in \cref{sec:supp_spatial_relationships}.

\subsection{Object Renderings}
\label{sec:supp_object_renderings}

\begin{figure}
    \centering
    \includegraphics[width=\linewidth]{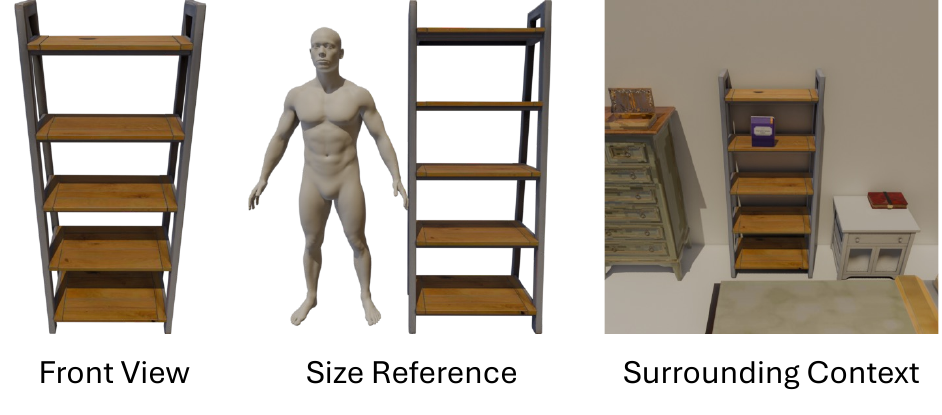}
    \vspace{-1.5em}
    \caption{
        Example renderings of the three object rendering types used in \ours: Front View, Size Reference, and Surrounding Context.
    }
    \label{fig:rendering_types}
    \vspace{-1.5em}
\end{figure}

In \ours, we use three types of object renderings across our metrics:
1) \textbf{Front View}: The object is positioned in the center of the image, zoomed in, and rendered from the front with no other objects visible.
2) \textbf{Size Reference}: The object is rendered with a 170 cm tall human figure on the left side for size reference.
3) \textbf{Surrounding Context}: The object is positioned in the center of the image and zoomed out to show the surrounding context.
These renderings help an LLM to understand the object's appearance, size, and context, respectively, for various evaluation tasks.
\cref{fig:rendering_types} shows example renderings of these three rendering types.

\subsection{Implementation Details}
\label{sec:supp_implementation_details}

We provide additional details for object-object relationship and object support metrics below.

\subsubsection{Object-Object Relationship}
For each object-object relationship in the annotations, we first map the annotated relationship into one or more of the 13 predefined spatial relationships (see \cref{sec:supp_spatial_relationships}) using an LLM.
After mapping, we locate all objects in the scene that match the categories specified in the relationship.
We consider all possible object combinations and compute a relationship score for each of them using the predefined spatial relationships.
All mapped relationships must be satisfied for an object combination to satisfy the original specification.

\subsubsection{Object Support}
To evaluate whether an object is stably supported, we first give two rendered images (front view and surrounding context) to an LLM and ask it to determine the support type of the object (one of: \texttt{ground}, \texttt{object}, \texttt{wall}, or \texttt{ceiling}).
Based on the type, we determine the object's support direction in its local frame (e.g., downward for \texttt{ground} and backward for \texttt{wall}) and cast rays towards that direction, from the object mesh vertices that are closest in that direction, and check for ray contacts with other geometries in the scene within 1 cm.
\texttt{Wall} and \texttt{ceiling} objects are considered supported if there are any valid contact points. (e.g., a ceiling lamp hanging from one point on the ceiling).
For \texttt{ground} and \texttt{object} types, we construct a convex hull from the contact points and project the object centroid in the gravity direction.
The object is considered supported if the projection is within the hull.
We repeat this process for all object instances and report the percentage of objects that are supported.

\subsection{Predefined Spatial Relationships}
\label{sec:supp_spatial_relationships}

\subsubsection{Object-Object Relationships}
Our object-object relationship metric uses a set of 13 predefined spatial relationships between objects.
We describe the implementation details of these relationships below.
Unless otherwise specified, we use a threshold of 0.5 to determine if a relationship is positive or negative.

\mypara{Inside and Outside} determine whether an object A is inside or outside another object B (e.g., a cup is inside a cabinet).
We sample points within object A's bounding box and compute a score based on the percentage of points that are inside object B's bounding box.

\mypara{Face} determines whether an object A is facing another object B (e.g., a sofa is facing a TV).
We sample points within object A's bounding box and shoot rays from these points in the direction of object A's front vector.
If there are no intersections with object B, the relationship is negative.
Otherwise, we take the mean coordinates of all intersection points and compute a score based on the angle between the front vector of object A and the vector from object A's centroid to the mean intersection point (ignoring the vertical axis).
The score is 1.0 if the angle is 0.0, and drops to 0.0 as the angle approaches 30.0 degrees.

\mypara{Side\_of} determines whether an object A is on one of the six sides (top, bottom, left, right, front, back) of another object B (e.g., a nightstand is on the left side of a bed).
We sample points within object A's bounding box and compute a score based on the percentage of points that are on the specific side of object B's bounding box (in object B's local coordinate frame), excluding points that are inside object B's bounding box.
Object B's bounding box is extended by 25\% in each dimension to account for slight misalignment.

\mypara{Side\_region} determines whether an object A is in one of the six side regions of another object B (e.g., a book is on the left side of a bookshelf).
The difference between this relationship and \texttt{side\_of} is that object A can be inside object B's bounding box.
The implementation is the same as \texttt{side\_of}, except that points inside object B's bounding box are not excluded and no extension is applied to object B's bounding box.

\mypara{Long\_short\_side} determines whether an object A is on the long or short side of another object B (e.g., a chair is on the long side of a table).
The long and short sides are determined based on object B's bounding box dimensions.
We sample points within object A's bounding box and compute a score based on the percentage of points that are on the long or short sides of object B.

\mypara{On\_top} determines whether an object A is on top of another object B (e.g., a book is on top of a table).
This relationship is specific for objects that are precisely placed on top of another object.
The implementation is the same as \texttt{side\_of}, with the top side of object B as the reference side, and no extension is applied to object B's bounding box.

\mypara{Middle\_of} determines whether an object A is in the middle of another object B (e.g., a pillow is in the middle of a bed).
We compute the distance between the centroids of object A and object B in 2D (ignoring the vertical axis) and apply a Gaussian with mean 0.0 and standard deviation 0.25 to compute the score.

\mypara{Surround} determines whether a group of objects $N$ surrounds a central object B (e.g., two chairs and two armchairs surround a table).
First, we calculate the ideal angle $A$ for uniformly distributing the objects in $N$ around object B as $A = \frac{2\pi}{|N|}$ and the mean distance $D$ between the centroids of objects in $N$ and object B.
Next, we compute the distance deviation $d_i$ and angle deviation $a_i$ from the ideal distance and angle for each object $i$ in $N$.
Each deviation is normalized by $D$ and $A$, respectively, and clipped to be within $[0, 1]$.
Finally, the score $s$ is computed as:
\begin{equation}
    s = \frac{1}{2|N|} \sum_{i=1}^{|N|} (1 - d_i)^2 + (1 - a_i)^2
\end{equation}

\mypara{Next\_to, Near, Across, and Far} determine whether an object A is within a certain distance from another object B (e.g., a TV is near a plant).
\texttt{next\_to} is defined as $0 \leq d \leq 0.5$, \texttt{near} is defined as $0.5 \leq d \leq 1.5$, \texttt{across} is defined as $1.5 \leq d \leq 4.0$, and \texttt{far} is defined as $d \geq 4.0$, where $d$ is the distance in meters between the closest points of the two objects.
The score is 1 if $d$ falls within the specified range, and drops as $d$ deviates from the range using a Gaussian with mean 0.0 and standard deviation 0.25.

\subsubsection{Object-Architecture Relationships}
Our object-architecture relationship metric uses a set of 10 predefined spatial relationships between objects and architecture.
We describe the implementation details of these relationships below.
Same as the object-object relationships, we use a threshold of 0.5 to determine if a relationship is positive or negative.

\mypara{Next\_to, Near, Across, and Far} are defined the same as in the object-object relationships.

\mypara{Inside\_room} determines whether an object A is inside a room (e.g., a chair is inside a living room).
We sample points within object A's bounding box and cast rays towards the room's floor plane.
The score is computed based on the percentage of points that intersect with the room's floor plane.

\mypara{Middle\_room} determines whether an object A is in the middle of a room (e.g., a rug is in the middle of a room).
We compute the distance between the centroid of object A and the room's centroid in 2D (ignoring the vertical axis) and apply a Gaussian with mean 0.0 and standard deviation $\frac{o}{2} + (1 - \frac{o}{r})$, where $o$ is the longer side of the object's 2D dimensions, and $r$ is the mean 2D room dimensions to compute the score, taking into account the object's size and the room's size.

\mypara{Corner\_room} determines whether an object A is in a corner of a room (e.g., a plant is in a corner of a room).
For every pair of walls in the room, we compute the distance scores between object A and the two walls similar to the \texttt{next\_to} relationship but with a range of $0 \leq d \leq 0.8$ in meters and a Gaussian with mean 0.0 and standard deviation 0.25.
We also compute the dot product between the front vectors of the two walls to determine if they are perpendicular.
If they are perpendicular, the score for this pair of walls is the product of the two distance scores.
The final score is the maximum score among all pairs of walls.

\mypara{On\_wall} determines whether an object A is on a wall (e.g., a painting is on a wall).
We first compute a score $s_f$ based on the percentage of points sampled within object A's bounding box that lie in front of the wall.
Next, we compute a score $s_d$ based on the closest distance between object A and the wall similar to the \texttt{next\_to} relationship but with a range of $0 \leq d \leq 0.01$ in meters and a Gaussian function of mean 0.0 and standard deviation 0.01.
The final score is the product of $s_f$ and $s_d$.

\mypara{Against\_wall} determines whether an object A is against a wall (e.g., a sofa is against a wall).
The implementation is the same as \texttt{on\_wall}, except that the range for $d$ is $0 \leq d \leq 0.3$ and the Gaussian function has a standard deviation of 0.1.

\mypara{Hang\_ceiling} determines whether an object A is hanging from the ceiling (e.g., a light is hanging from the ceiling).
The implementation is similar to \texttt{next\_to}, except that the reference element is the ceiling, the range for $d$ is $0 \leq d \leq 0.01$, and the Gaussian function has a standard deviation of 0.03.

\section{\ours with Open-Source VLM}
\label{sec:supp_qwen}

To assess the generality of our framework, we re-ran SceneEval on 500 scenes generated from the 100 manually written descriptions using Qwen2.5-VL-7B-Instruct~\cite{bai2025qwen2}, a publicly available open-source VLM that we were able to run locally with available resources.

The overall evaluation trends remain consistent with those obtained using our original model (GPT-4o~\cite{achiam2023gpt}).
Agreement with our manual evaluation across the four fidelity metrics is 80.65\% (Object Count), 76.64\% (Attribute), 91.11\% (Object-Object Relationship), and 87.72\% (Object-Architecture Relationship), with Cohen’s kappas of 0.50, 0.26, 0.43, and 0.47, respectively.
For the user study, we report agreement using only the subset of user-rated scenes that overlap with the 100 manual descriptions. On this subset, agreement is 77.55\%, 75.00\%, 78.67\%, and 83.48\%, with Cohen’s kappas of 0.48, 0.12, 0.32, and 0.48, respectively.

These results show that while agreement scores using Qwen2.5-VL-7B-Instruct are lower than those achieved with GPT-4o (a significantly larger and more capable model) --- particularly for the attribute metric, which solely relies on the VLM for evaluation --- they still follow the same general trends and remain within a reasonable range.
This highlights the benefit of using stronger VLMs for specific perception components (such as attribute recognition), while also demonstrating that SceneEval’s evaluation pipeline is modular and not overly dependent on any single model.

\section{Limitations}
\label{sec:supp_limitations}

While our dataset and metrics provide a better coverage of important aspects of scene generation compared to existing metrics, they are not perfect.
We provide a discussion of the limitations of \ours, \ourdataset, and our semi-automatic data generation process below.

\subsection{Limitations in \ours}
\label{sec:supp_limitations_ours}

First, our metrics currently do not consider whether objects in the generated scenes are placed according to ``common sense'' expectations, even if they are not explicitly specified in the input text descriptions.
For example, large furniture items, like bookshelves, are typically placed against walls, except when they are used to divide spaces.
Such common sense expectations are crucial for realism of the generated scenes and is an important aspect for evaluating scene generation models.
Unfortunately, such expectations are less well-defined.
As a result, incorporating them into the evaluation metrics is challenging and requires further research.

Second, \ours's execution time currently scales with the number of objects in the scene.
As scenes get more complex, the time required to perform object matching and compute the metrics also increases, as there are more objects to process.
Exploring parallelization and other optimization techniques to reduce the execution time is an important direction for future work.

\subsection{Limitations in \ourdataset}
\label{sec:supp_limitations_ourdataset}

While our dataset includes a broader range of room types than prior work, its scale remains limited compared to datasets in other domains (e.g., image), which often contain thousands to millions of entries.
Expanding the dataset further, especially through scalable automatic methods, would allow for more comprehensive evaluations and a deeper analysis of model capabilities.

Additionally, the authors who created the initial 100 descriptions and annotations, and who validated and corrected the semi-automatically generated ones, are primarily from Asia and have lived in Western countries. None are trained interior designers or architects. As a result, the dataset may reflect cultural assumptions and expectations that are not universally representative.

\subsection{Semi-Automatic Data Generation Limitations}
\label{sec:supp_limitations_generation}

Despite the advantages of semi-automatic data generation over manually writing descriptions and annotations, it is not without its challenges.
During data generation, we encountered four main failure modes that limit scalability and prevent fully automatic generation:

\mypara{Missing annotations.}
The most common issue is the omission of object attributes and object-object relationships.
Specifically, 101 missing attributes (e.g., ``wooden'') were identified and manually added during the generation of 210 medium and hard scene descriptions.
Additionally, we manually added 154 missing object-object relationships. 
Many cases involved the omission of one among multiple constraints for an object; for instance, given the description ``On the desk, a pen is right of a laptop,'' the annotation ``pen on desk'' was missing.
Such omissions are especially common in hard scenes, whose descriptions are longer and contain more attributes and relations.

\mypara{Hallucination.}
During consecutive generation with a lengthy conversation history, the VLM often erroneously includes annotations from previously generated descriptions.
This issue becomes more frequent after generating more than five descriptions and their annotations.
It also affects the anchor index field, where the VLM may output incorrect values (e.g., specifying an index of 14 when only two objects are involved in the relationship).

\mypara{Inconsistency in object attributes.}
The VLM occasionally merges multiple attributes into a single non-atomic entry.
For example, it may produce ``round wooden'' as one attribute instead of ``round'' and ``wooden''.
This violates our expectation that each attribute is independent, and requires manual splitting during validation.

\mypara{Decreasing diversity.}
In addition to increased hallucination, we observed that with longer generation history, the VLM tends to produce scene descriptions with reduced diversity in object selection or spatial arrangement, requiring manual intervention such as resetting the conversation history or prompting with specific scene ideas.

These limitations currently prevent fully automatic dataset generation. 
Improving the generation process remains an important direction for future work toward building larger and more diverse datasets with reduced manual effort.

\section{User Study Details}
\label{sec:supp_user_study}

\begin{figure*}
    \centering
    \includegraphics[width=\linewidth]{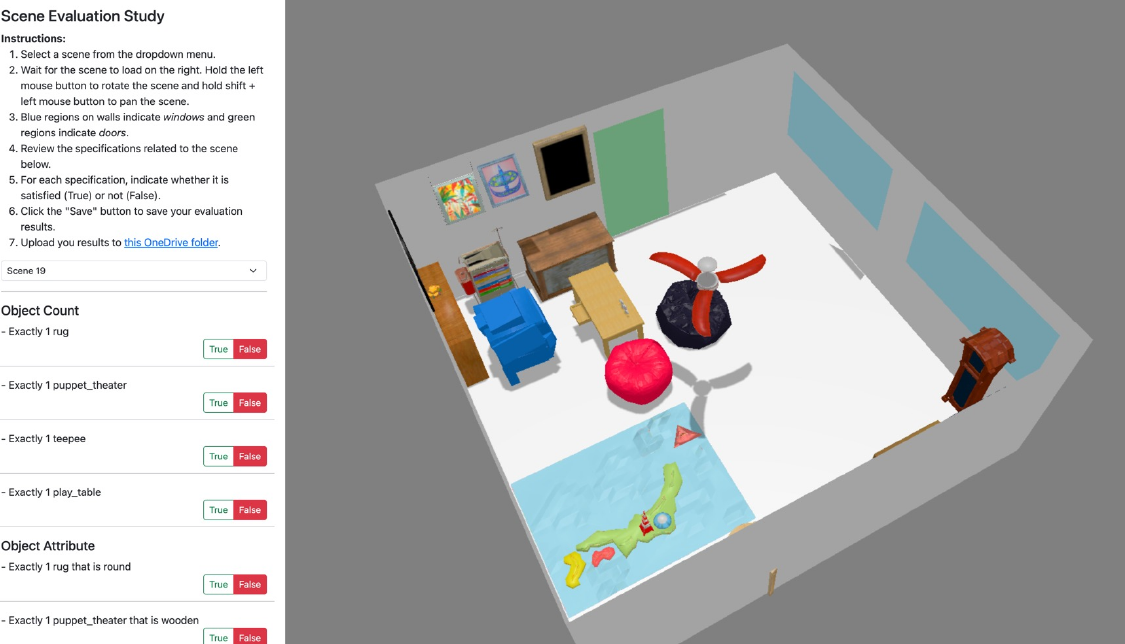}
    \vspace{-1.5em}
    \caption{
        Interface used for the user study.
        \textbf{Top left:} Study instructions and interface guidelines.
        \textbf{Bottom left:} Scene properties to evaluate.
        \textbf{Right:} Interactive 3D viewer with free camera control for detailed inspection.
    }
    \label{fig:user_study_ui}
    \vspace{-1em}
\end{figure*}

\cref{fig:user_study_ui} shows the interface used in our user study.
Participants are presented with a 3D scene in an interactive viewer, enabling them to freely inspect specific details while evaluating the listed scene properties.
Participants are instructed to carefully examine both the scene and the expected properties, selecting \textit{True} or \textit{False} to indicate whether each property is satisfied.
Once complete, they save their responses to a dedicated cloud storage location.

\section{VLM Prompts}
\label{sec:supp_llm_prompts}

\ours uses an VLM to assist in parts of the evaluation framework.
We provide the system prompt in \cref{sec:supp_llm_prompts_system_prompt} and the evaluation task prompts in \cref{sec:supp_llm_prompts_task_prompts}.
Additionally, we provide the prompts for semi-automatic data generation in \cref{sec:supp_llm_prompts_data_gen}.

\subsection{System Prompt}
\label{sec:supp_llm_prompts_system_prompt}
The system prompt provides the VLM with the overall context about the tasks and the role it plays in the evaluation framework.

\begin{minted}[breaklines, breakafter=d, escapeinside=||, fontsize=\scriptsize]{yaml}
system: >
  You are an expert in interior design.
  You have seen thousands of interior designs and have a good understanding of the spatial arrangement of objects in a room.
  Now, you are working as an evaluator for a design company.
  Use your expertise in interior design to evaluate the spatial arrangement of objects in the given scene according to the task instructions.
  When you are required to include object descriptions in your response, respond exactly as they are provided in the task instructions word for word.
  When you are required to give a specific side for a response to a relationship, use only the sides provided in the task instructions.
\end{minted}

\subsection{Evaluation Task Prompts}
\label{sec:supp_llm_prompts_task_prompts}
There are six tasks in \ours that use an VLM for assistance.
We provide the prompts for each task below.

\subsubsection{Object Matching}
This task asks the VLM to match objects in the scene to the object categories specified in the ground truth annotation.
Given a front-view image of an object and the object categories, the VLM is asked to determine if the object belongs to any of the specified categories and provide a justification.

\begin{minted}[breaklines, breakafter=d, fontsize=\scriptsize]{yaml}
obj_matching: >
  The user specified the scene to contain objects of certain categories.
  To facilitate further evaluation, you need to match the objects in the scene to the object categories specified by the user.
  You are provided an image of one of the objects in the scene.
  Does the object in the image belong to any of the object categories specified by the user?
  Respond in the given response schema. Here are two example responses:
  ```
  provided_categories: ["chair", "table", "lamp"]
  matched: True
  matched_category: "chair"
  reason: "The object in the image is a chair."
  ```
  ```
  provided_categories: ["chair", "table", "lamp"]
  matched: False
  matched_category: ""
  reason: "The object in the image is a sofa, which does not match any of the specified categories."
  ```
  If the object in the image does not belong to any of the object categories specified by the user, respond with "matched: False" and "matched_category: "".
  Here is the list of object categories that the user specified to match against:
  "<TARGET_CATEGORIES>"
\end{minted}

\subsubsection{Object Attribute}
This task asks the VLM to determine if the objects in the scene satisfy the attribute requirements in the annotations.
For each object of interest, the VLM is provided with two images: one from the front view and one with a human model on the side for scale.
The VLM is asked to determine if the object satisfies the attribute requirements and provide a reason for its decision.

\begin{minted}[breaklines, breakafter=d, fontsize=\scriptsize]{yaml}
obj_attribute: >
  The user specified the scene to contain objects with certain attributes.
  You are provided with images of instances of objects in the scene with the same category.
  There are two images for each object instance: one from the front view and one with a 170cm human model for scale.
  The images are in the following order: obj1_front, obj1_scale, obj2_front, obj2_scale, ...
  Given these images, how many of these objects satisfy the attribute requirements specified by the user?
  Note that the human model is included in the images solely for scale reference and should not be considered as part of the evaluation.
  Respond in the given response schema. Here is an example response:
  ```
  category: "chair",
  num_instances: 3,
  [
    {
      "instance": 0,
      "attribute": "red",
      "satisfied": True,
      "reason": "This chair is red."
    },
    {
      "instance": 1,
      "attribute": "red",
      "satisfied": False,
      "reason": "This chair is blue."
    },
    ...
  ]
  ```
  The attribute requirements are as follows:
  "<OBJ_ATTRIBUTES>"
  Here are the renderings of "<OBJ_COUNT>" instances of object with category "<OBJ_CATEGORY>" in the scene.
\end{minted}

\subsubsection{Object Support Type}
This task asks the VLM to identify the support type of objects in the scene.
For each object, the VLM is provided with two images: one from the front view and one slightly zoomed out to show the surrounding area.
The VLM is asked to pick the support type of the object from the predefined types and provide a reason for its decision.

\begin{minted}[breaklines, breakafter=d, fontsize=\scriptsize]{yaml}
obj_support_type: >
  Objects in the scene are placed on the ground, on wall, on ceiling, or on other objects.
  You are given two images of an object in the scene: one from the front view and one slightly zoomed out to show the surrounding area.
  Using the images, identify the support type of the object.
  The support type of an object is the surface on which the object is placed.
  Here are the support types for objects:
  - ground: The object is placed on the ground. (e.g., "table on the ground")
  - wall: The object is placed on the wall. (e.g., "painting on the wall")
  - ceiling: The object is placed on the ceiling. (e.g., "lamp hanging from the ceiling")
  - object: The object is placed on a surface of another object. (e.g., "book on the table")
  Respond in the given response schema. Here is two example responses:
  ```
  support_type: "ground",
  reason: "The table is placed on the ground."
  ```
  ```
  support_type: "wall",
  reason: "The painting is placed on the wall."
  ```
  If the object appears to be a ceiling light, carefully consider the image as it may be difficult to see that the object is hanging from the ceiling.
\end{minted}

\subsubsection{Object Functional Sides}
This task asks the VLM to identify the functional sides of objects in the scene.
The functional sides of an object are the sides that need to be accessible for the object to be used properly.
The VLM is provided with descriptions of the objects in the scene and asked to identify the functional sides of each object with a justification.

\begin{minted}[breaklines, breakafter=d, fontsize=\scriptsize]{yaml}
obj_functional_sides: >
  Objects in the scene have functional sides that are important for their placement and use.
  The functional sides of an object are the sides that need to be accessible for the object to be used properly.
  Here, only consider these four sides of an object: ["front", "back", "left", "right"].
  If an object is placed in a scene, at least one of its functional sides should be accessible for the object to be considered properly placed.
  If an object has multiple functional sides, this means that the object can be used from any of these sides and there is no difference in importance between them.
  Otherwise, only consider the most important functional side as the sole functional side of the object.
  Here are some examples of different cases:
  - Objects that have equal importance for their functional sides:
    - bed: ["front", "left", "right"]
    - dining_table: ["front", "back", "left", "right"]
  - Objects that have a significant front side:
    - desk: ["front"]
    - sofa: ["front"]
  - Objects that can be moved so all sides are functional:
    - dining_chair: ["front", "back", "left", "right"]
    - stool: ["front", "back", "left", "right"]
  You are provided with descriptions of the objects in the scene.
  The task is to identify the functional sides of each of the objects.
  Note that for small objects like cups and books that are placed on a surface, do not consider their functional sides and respond with an empty list.
  Respond in the given response schema. Here is an example response:
  ```
  [
    {
      "obj_description": "bed.n.01 - bed description",
      "functonal_sides": ["front", "left", "right"],
      "reason": "These three sides of a bed have equal importance for accessibility and as long as one of them is accessible, the bed is considered properly placed."
    },
    {
      "obj_description": "chair.n.01 - chair description",
      "functonal_sides": ["front", "back", "left", "right"],
      "reason": "All four sides of a chair are functional because it can be moved and used from any side."
    },
    {
      "obj_description": "cup.n.01 - cup description",
      "functonal_sides": [],
      "reason": "Cups are small objects that do not have functional sides."
    }
    ...
  ]
  ```
  The descriptions of the objects in the scene are as follows:
  "<OBJ_DESCRIPTIONS>"
\end{minted}

\subsubsection{Object Relationship Mapping}
This task asks the VLM to map open-vocabulary object-object relationships in the annotations to predefined spatial relationship types.
For each input relationship, the VLM can choose multiple relationship types if multiple types are required to fully describe the relationship.
The prompt provides the definition of the predefined spatial relationship types, with examples and guidelines for mapping the relationships.
The VLM is asked to provide the mapped relationship types for each input relationship along with the necessary information.

\begin{minted}[breaklines, breakafter=d, fontsize=\scriptsize]{yaml}
obj_relationship_mapping: >
  The user specified the scene to contain certain relationships between objects.
  An object-object relationship is a spatial relationship between two or more objects in the scene.
  In which, an anchor object is the object that is used as a reference point to compare against.
  Here are some examples to illustrate the concept of anchor object:
  - "chair next to the table": the table is the anchor object.
  - "lamp near the sofa": the sofa is the anchor object.
  You are provided with manually annotated relationships between objects in the scene.
  The task is to map the mentioned relationships into one or more of the predefined spatial relationship type here:
  - inside_of: The target object is inside the anchor object. (e.g., "cup inside the cabinet")
  - outside_of: The target object is outside the anchor object. (e.g., "toy outside the box")
  - face_to: The target object is facing the anchor object. (e.g., "sofa facing the TV")
  - side_of: The target object is at one of the six sides (left, right, front, back, top, bottom) of the anchor object. (e.g., "nightstand left of the bed")
  - side_region: The target object is inside the anchor object at one of the six sides (left, right, front, back, top, bottom). (e.g., "book on the left side of the shelf", "phone on the left side of the table")
  - long_short_side_of: The target object is specifically at a long or short side of the anchor object. (e.g., "book at the long side of the table")
  - on_top: The target object is on top of the anchor object at its top-most surface and not inside it. (e.g., "book on top of the table", but not applicable for "book on a "bookshelf" because the book is technically inside the bookshelf - use inside_of instead)
  - middle_of: The target object is in the middle of the anchor object. (This only compares the objects in 2D, e.g., "pillow in the middle of the bed")
  - surround: Multiple target objects (can be different types) are circled around one anchor object. (e.g., "four chairs surrounding the table")
  - next_to: The target object is next to the anchor object within 0 to 0.5m (e.g., "chair next to the table")
  - near: The target object is near the anchor object within 0.5 to 1.5m. (e.g., "sofa near the TV")
  - across_from: The target object is far from the anchor object within 1.5 to 4m. (e.g., "lamp across the room from the sofa")
  - far: The target object is far from the anchor object beyond 4m. (e.g., "painting far from the bed")
  - None: None of the predefined spatial relationships above match the relationship
  You can choose multiple relationship types for a single input relationship if it requires multiple types to fully describe the relationship.
  Here is an example of relationships that require multiple types to fully describe them:
  - "table at the foot of the bed" needs both "side_of" and "next_to" relationship types to fully describe it.
  Here are some additional guidelines for mapping the relationships:
  - When choosing side_of and side_region, you must also specify the side of the anchor object (left, right, front, back, top, bottom).
  - When choosing long_short_side_of, you must also specify the side of the anchor object (long, short).
  - For side ambiguous relationships, like "next to" or "adjacent to", simply choose the distance-based relationship (next_to, near, across_from, far).
  - When you choose multiple types for a single relationship, and some of the types require specifying a side, you must specify the side for all types in the same order as the types are listed.
    - Use "None" for the side when a type does not require specifying a side.
    - For example, if you choose both "side_of" and "next_to" for a relationship, you must specify the sides as ["front", None].
  - Even if the relationship type does not require specifying a side, you must still provide a side as "None" in the response at the corresponding index.
  - When the anchor object is not specified (i.e., when the anchor index is -1), put the first object in the relationship as the anchor object in your response.
  - The other_object_counts are the number of objects that are part of the relationship for each object category in other_objects in the same order.
  - When none of the predefined spatial relationships match the relationship, put "None" as the relationship type and provide a reason. (Do not put an empty list.)
  Respond in the given response schema. Here is an example response:
  ```
  [
    {
      "relationship": "beneath - objects: box, bed, with the object with index: 0 being the anchor",
      "anchor_object": "bed",
      "other_objects": ["box"],
      "other_object_counts": [1],
      "relationship_types": ["side_region"],
      "sides": ["bottom"],
      "reason": "Box beneath the bed is considered as the box being inside the bed at the bottom side."
    },
    {
      "relationship": "next_to - objects: lamp, chair, with the object with index: 0 being the anchor",
      "anchor_object": "lamp",
      "other_objects": ["chair"],
      "other_object_counts": [1],
      "relationship_types": ["next_to"],
      "sides": [None],
      "reason": "Chair next to the lamp is considered as the chair being next to the lamp."
    },
    {
      "relationship": "at the foot of - objects: bed, table, with the object with index: 0 being the anchor",
      "anchor_object": "bed",
      "other_objects": ["table"],
      "other_object_counts": [1],
      "relationship_types": ["side_of", "next_to"],
      "sides": ["front", None],
      "reason": "Table at the foot of the bed is considered as the table being at the front side of the bed and next to it."
    },
    {
      "relationship": "surround - objects: table, chair:0, chair:1, chair:2, sofa, with the object with index: 0 being the anchor",
      "anchor_object": "table",
      "other_objects": ["chair", "sofa"],
      "other_object_counts": [3, 1],
      "relationship_type": ["surround"],
      "side": [None],
      "reason": "Three chairs and a sofa surrounding the table is considered as the chairs and the sofa surrounding the table."
    },
    {
      "relationship": "diagnoally across - objects: table, chair, with the object with index: 0 being the anchor",
      "anchor_object": "table",
      "other_objects": ["chair"],
      "other_object_counts": [1],
      "relationship_types": None
      "sides": [None],
      "reason": "No appropriate relationship type found for this relationship - chair diagonally across the table."
    }
    ...
  ]
  ```
  The annotated relationships between objects in the scene are as follows:
  "<RELATIONSHIPS>"
\end{minted}

\subsubsection{Architectural Relationship Mapping}
This task asks the VLM to map open-vocabulary relationships between objects and architectural elements in the annotations to predefined spatial relationship types.
Similar to the object relationship mapping task, the VLM is provided with the definitions of the predefined spatial relationship types, with examples and guidelines for mapping the relationships.
The VLM is asked to provide the mapped relationship types for each input relationship along with the necessary information.

\begin{minted}[breaklines, breakafter=d, fontsize=\scriptsize]{yaml}
arch_relationship_mapping: >
  The user specified the scene to contain certain relationships between objects and architectural elements.
  An architectural element is a structural component of a building, such as a wall, floor, ceiling, or room.
  Here are some examples of relationships between objects and architectural elements:
  - "painting on the wall"
  - "bookshelf against the wall"
  You are provided with manually annotated relationships between objects and architectural elements in the scene.
  The task is to map the mentioned relationships into one of the predefined spatial relationship type:
  - inside_room: The target object is inside the room. (e.g., "sofa inside the room")
  - middle_of_room: The target object is in the middle of the room. (e.g., "table in the middle of the room")
  - next_to: The target object is next to an architectural element within 0 to 0.5m. (e.g., "chair next to the wall")
  - near: The target object is near an architectural element within 0.5 to 1.5m. (e.g., "lamp near the door")
  - across_from: The target object is far from an architectural element within 1.5 to 4m. (e.g., "art across from the wall")
  - far: The target object is far from an architectural element beyond 4m. (e.g., "table far from the window")
  - on_wall: The target object is on the wall (must be directly in front of the wall). (e.g., "painting on the wall")
  - against_wall: The target object is against the wall (must be directly in front of the wall). (e.g., "bookshelf against the wall")
  - corner_of_room: The target object is at the corner of the room. (e.g., "chair at the corner of the room")
  - hang_from_ceiling: The target object is hanging from the ceiling. (e.g., "lamp hanging from the ceiling")
  - None: None of the predefined spatial relationships above match the relationship
  Here are some additional guidelines for mapping the relationships:
  - When specifying the architectural element type, select from the following: ["wall", "floor", "ceiling", "room", "window", "door"].
  - When choosing floor or room, you must also specify the specific floors from the provided list of floor IDs.
    - If the IDs are not informative enough and you cannot determine the specific floor, choose all floors in the scene.
    - If the relationship is not specific to a floor, choose all floors in the scene.
  - When none of the predefined spatial relationships match the relationship, put "None" as the relationship type and provide a reason.
  Respond in the given response schema. Here is an example response:
  ```
  [
    {
      "relationship": "on - object: painting, with respect to architectural element: wall"
      "target_object": "painting",
      "architectural_element_type": "wall",
      "relationship_type": "on_wall",
      "specific_floors": [],
      "reason": "The painting is on the wall."
    },
    {
      "relationship": "along - object: bookshelf, with respect to architectural element: wall"
      "target_object": "bookshelf",
      "architectural_element_type": "wall",
      "relationship_type": "against_wall",
      "specific_floors": [],
      "reason": "The bookshelf is along a wall means that it is in front of and against the wall."
    },
    {
      "relationship": "corner - object: chair, with respect to architectural element: bedroom"
      "target_object": "chair",
      "arch_element_type": "room",
      "relationship_type": "corner_of_room",
      "specific_floors": ["floor_bedroom_001", ...]
      "reason": "The chair is at the corner of the room."
    }
    ...
  ]
  ```
  The annotated relationships between objects and architectural elements in the scene are as follows:
  "<RELATIONSHIPS>"
  Here are all the floors in the scene that you can choose from:
  "<FLOOR_IDS>"
\end{minted}

\subsection{Semi-Automatic Data Generation}
\label{sec:supp_llm_prompts_data_gen}

The semi-automatic data generation process uses a VLM to generate scenes descriptions and annotations given a set of manually created entries as in-content examples.

\subsubsection{System Prompt}
The system prompt provides the VLM with the task description, the annotation schema, and other relevant guidelines.
\begin{minted}[breaklines, breakafter=d, escapeinside=||, fontsize=\scriptsize]{yaml}
system: >
  You are tasked to generate annotations for scenes given a text description of the scene.
  There are four types of annotations that you need to generate:
  - Object Count: The number of objects in the scene. The schema is:
    - {eq, lt, gt, le, ge},<instance_count>,<obj_reference>
    - e.g., "eq,3,chair" means that there are exactly 3 chairs in the scene
    - <obj_reference> is the category name of the object, e.g., "chair", "table", "lamp", and not the object instance name
      - The object category must not include a descriptive attribute
      - e.g., "fridge" instead of "mini_fridge", as "mini" can be captured in the object attributes
      - On the other hand, "floor_lamp" is acceptable as it is a specific type of lamp and because "floor" is not an attribute
  - Object Attribute: The attributes of the objects in the scene. The schema is:
    - {eq, lt, gt, le, ge},<instance_count>,<obj_reference>,<attribute>
    - e.g., "eq,1,chair,red" means that there is exactly 1 red chair in the scene
    - All <obj_reference> must refer to an object category that is mentioned in the object count for the same scene
  - Object-Object Relationship: The relationships between objects in the scene. The schema is:
    - {eq, lt, gt, le, ge},<instance_count>,<relationship>, <anchor_index>,<obj_reference_0>, <obj_reference_1>,<obj_reference_n>
    - e.g., "eq,1,front,0,desk,chair" means that the object at index 0 (desk) is the anchor object and a chair is in front of it, and there is exactly 1 such relationship in the scene
    - <obj_refernce> is category name of the object, e.g., "chair", "table", "lamp", and not the object instance name
    - All <obj_reference> must refer to an object category that is mentioned in the object count for the same scene
    - Note that the <relatioship> must be broken into pairwise relationships, except for "surround" as in "three chairs surrounding the table"
      - e.g., "two chairs next to the table" should be broken into as "eq,2,next,0,table,chair"
      - e.g., "two nightstands on two sides of the bed" should be broken into two relationships: "eq,1,left,0,bed,nightstand" and "eq,1,right,0,bed,nightstand"
    - The <anchor_index> is the index of the anchor object in the scene description where the corresponding object is the reference point for the relationship
      - The index starts from 0 and is based on the order of the objects in the scene description
      - The anchor object can be any object in the scene
      - e.g., in the relationship "eq,1,left,0,bed,nightstand", the anchor object is the bed (index 0) and the nightstand is to the left of it
      - e.g., in the relationship "eq,1,face,0,desk,chair", the anchor object is the desk (index 0) and the chair is facing it
      - Fix the anchor index to be 0 and instead re-arrange the obj_references as needed
  - Object-Architecture Relationship: The relationships between objects and architectural elements in the scene. The schema is:
    - {eq, lt, gt, le, ge},<instance_count>,<relationship>, <obj_reference>,<arch_reference>
    - e.g., "eq,1,against,bookshelf,wall" means that the bookshelf is against the wall, and there is exactly 1 such relationship in the scene
    - <obj_reference> is the category name of the object, e.g., "chair", "table", "lamp", and not the object instance name
    - All <obj_reference> must refer to an object category that is mentioned in the object count for the same scene
    - <arch_reference> can be one of the following: ["wall", "floor", "ceiling", "room", "window", "door"] or a specific room type, e.g., "bedroom", "living_room", "kitchen", etc
      - It cannot be a specific instance of a category, e.g., "wall_1", "floor_2", "ceiling_3", "room_4", or "kitchen_5", etc
    - You do not need to specify something is on the floor as it is expected and adding one for each object is redundant
  The difference between object-object relationships and object-architecture relationships is that the former is between two objects (or more if "surround"), while the latter is between an object and an architectural element.
  For example, door, window, wall, floor, and ceiling are architectural elements and any relationship with them must be an object-architecture relationship.
  Otherwise, if the relationship only involves objects, it is an object-object relationship.
  {eq, lt, gt, le, ge} are the quantifier operators for equal, less than, greater than, less than or equal to, and greater than or equal to respectively.
\end{minted}

\subsubsection{In-Content Example}
This prompt provides the VLM with a set of scene descriptions and their corresponding annotations as in-content examples.
\begin{minted}[breaklines, breakafter=d, escapeinside=||, fontsize=\scriptsize]{yaml}
show_examples: >
  Here are <NUM_EXAMPLES> manually annotated examples of the scene descriptions and their corresponding annotations:
  <IN_CONTEXT_EXAMPLES>
  Note that the annotations are in the same format as described above.
  The annotations are exhaustive and cover all the objects in the scene.
  Observe and learn from the examples, I will then ask you to generate annotations for a new scene description.
\end{minted}

\subsubsection{Generate Scene Description}
This task asks the VLM to generate scene descriptions.
There are two versions of the prompt: one for generating a single scene description and one for generating multiple scene descriptions.
\begin{minted}[breaklines, breakafter=d, escapeinside=||, fontsize=\scriptsize]{yaml}
generate_description: >
  Let's start a new cycle of generating scene descriptions and annotations.
  Your first task is to generate a scene description.
  There are three difficulty levels: easy, medium, and hard.
  - Easy:
    - Total number of all objects: <= 4
    - At most 4 large furniture objects (e.g., bed, sofa, table, chair)
    - Zero small objects (e.g., lamp, vase, book)
  - Medium:
    - Total number of all objects: 5 to 8
    - 0 to 3 small objects (e.g., lamp, vase, book)
    - Remaining objects are large furniture objects (e.g., bed, sofa, table, chair)
  - Hard:
    - Total number of all objects: >= 9
    - No limit on the number of small objects (e.g., lamp, vase, book)
  The examples you have seen are of the same difficulty level as the one you are going to generate.
  Use your knowledge of the world and the examples you have seen to generate a scene description.
  But you do not need to base your scene description on the examples you have seen.
  Be creative, you do not need to follow the examples.
  Avoid viewpoint-dependent descriptions like left wall, right wall, etc as these are not meaningful in the context of a 3D environment.
  Only use ASCII characters in the scene description, no special characters.
  Have variety in the way you write. It can be started with "A" or "There is" at first but you should move away from them as you generate more descriptions.
  Use different sentence structures and avoid repetitive phrases.
  While having variety, you should also maintain clarity.
  Describe object styles and how they are arranged in the scene - object-object relationships and object-architecture relationships, as needed.
  Here are specific instructions from the user:
  --- Begin user instructions ---
  <INSTRUCTION>
  --- End user instructions ---
  Now, generate a <DIFFICULTY> scene description for the user.
  Respond in the given response schema. Here is an example response:
  ```
  {
    difficulty_level: "easy", (copy the difficulty level from the user)
    instruction: "Generate a simple bedroom", (copy the instruction from the user)
    generated_scene_description: "A bedroom with a red bed, and a nightstand on the left side of the bed with a wardrobe in the corner",
    total_num_objects: 3,
    num_large_objects: 3,
    num_small_objects: 0,
    reason: (explain how you generated the scene description)
  }
  ```
batch_generate_descriptions: >
  Let's start a new cycle of generating scene descriptions and annotations.
  Your first task is to generate some scene descriptions.
  There are three difficulty levels: easy, medium, and hard.
  - Easy:
    - Total number of all objects: <= 4
    - At most 4 large furniture objects (e.g., bed, sofa, table, chair)
    - Zero small objects (e.g., lamp, vase, book)
  - Medium:
    - Total number of all objects: 5 to 8
    - 0 to 3 small objects (e.g., lamp, vase, book)
    - Remaining objects are large furniture objects (e.g., bed, sofa, table, chair)
  - Hard:
    - Total number of all objects: >= 9
    - No limit on the number of small objects (e.g., lamp, vase, book)
  The examples you have seen are of the same difficulty level as the ones you are going to generate.
  Use your knowledge of the world and the examples you have seen to generate scene descriptions.
  But you do not need to base your scene descriptions on the examples you have seen.
  Be creative, you do not need to follow the examples.
  Avoid viewpoint-dependent descriptions like left wall, right wall, etc as these are not meaningful in the context of a 3D environment.
  Only use ASCII characters in the scene description, no special characters.
  Have variety in the way you write. It can be started with "A" or "There is" at first but you should move away from them as you generate more descriptions.
  Use different sentence structures and avoid repetitive phrases.
  While having variety, you should also maintain clarity.
  Describe object styles and how they are arranged in the scene - object-object relationships and object-architecture relationships, as needed.
  Here are specific instructions from the user:
  --- Begin user instructions ---
  <INSTRUCTION>
  --- End user instructions ---
  Now, generate <NUM_DESCRIPTIONS> <DIFFICULTY> scene descriptions for the user.
  Respond in the given response schema. Here is an example response:
  ```
  [
    {
      difficulty_level: "easy", (copy the difficulty level from the user)
      instruction: "Generate a simple bedroom", (copy the instruction from the user)
      generated_scene_description: "A bedroom with a red bed, and a nightstand on the left side of the bed with a wardrobe in the corner",
      total_num_objects: 3,
      num_large_objects: 3,
      num_small_objects: 0,
      reason: (explain how you generated the scene description)
    },
    ...
  ]
  ```
\end{minted}

\subsubsection{Generate Annotations}
This task asks the VLM to generate annotations for the scene descriptions.
There are two versions of the prompt: one for generating annotations for a single scene description and one for generating annotations for multiple scene descriptions.
\begin{minted}[breaklines, breakafter=d, escapeinside=||, fontsize=\scriptsize]{yaml}
generate_annotations: >
  Now, you are going to generate annotations for the scene description you have just generated.
  Use what you have learned from the in-context examples.
  The user may have edited the scene description, so here is the final scene description:
  --- Begin scene description ---
  <SCENE_DESCRIPTION>
  --- End scene description ---
  Read the scene description carefully and generate annotations for it.
  Before generating the annotations, go over the definitions of the four types of annotations again.
  After that, go over the examples again and understand how the annotations should be generated.
  The annotations should be exhaustive and cover all the objects in the scene.
  The annotations should be in the same format as described above.
  Use underscore instead of space or hyphen in all annotations.
  If the description mentions absence of an object, you should add an annotation for it (e.g., "eq,0,chair").
  Note that if something is facing another object, not only can you use "front", you can also say "facing".
  Respond in the given response schema. Here is an example response:
  ```
  {
    scene_description: "a bedroom with a red bed, and a nightstand on the left side of the bed with a wardrobe in the corner",
    obj_counts: ["eq,1,bed", "eq,1,nightstand"],
    obj_attributes: ["eq,1,red,bed"],
    obj_obj_relationships: ["eq,1,left,0,bed,nightstand"],
    obj_arch_relationships: ["eq,1,corner,wardrobe,room"],
    reason: (explain how you generated the annotations)
  }
  ```
batch_generate_annotations: >
  Now, you are going to generate annotations for a set of scene descriptions you have just generated.
  Use what you have learned from the in-context examples.
  The user may have edited the scene descriptions, so here are the final scene descriptions:
  --- Begin all scene description ---
  <SCENE_DESCRIPTIONS>
  --- End all scene description ---
  Read the scene descriptions carefully and generate annotations for each of them.
  Before generating the annotations, go over the definitions of the four types of annotations again.
  After that, go over the examples again and understand how the annotations should be generated.
  The annotations should be exhaustive and cover all the objects in the scenes.
  The annotations should be in the same format as described above.
  Use underscore instead of space or hyphen in all annotations.
  If the description mentions absence of an object, you should add an annotation for it (e.g., "eq,0,chair").
  Note that if something is facing another object, not only can you use "front", you can also say "facing".
  Order the annotations to be consistent with the order of the scene descriptions.
  Respond in the given response schema. Here is an example response:
  ```
  [
    {
      scene_description: "a bedroom with a red bed, and a nightstand on the left side of the bed with a wardrobe in the corner",
      obj_counts: ["eq,1,bed", "eq,1,nightstand"],
      obj_attributes: ["eq,1,red,bed"],
      obj_obj_relationships: ["eq,1,left,0,bed,nightstand"],
      obj_arch_relationships: ["eq,1,corner,wardrobe,room"],
      reason: (explain how you generated the annotations)
    },
    ...
  ]
  ```
\end{minted}

\section{Scientific Artifacts}
\label{sec:supp_scientific_artifacts}

The licenses for the datasets and code used in this work are as follows.
For datasets,
3D-FRONT~\cite{fu20213dfront} and 3D-FUTURE~\cite{fu20213dfuture} are available under their respective terms of use\footnote{\href{https://gw.alicdn.com/bao/uploaded/TB1ZJUfK.z1gK0jSZLeXXb9kVXa.pdf?spm=a1z3i.a4.0.0.3f5beb1digOegr&file=TB1ZJUfK.z1gK0jSZLeXXb9kVXa.pdf}{3D-FRONT Terms of Use}}\textsuperscript{,}\footnote{\href{https://terms.aliyun.com/legal-agreement/terms/suit_bu1_ali_cloud/suit_bu1_ali_cloud202004171628_60052.html?spm=5176.14208604.0.0.a3c33cf7X7NfGY}{3D-FUTURE Terms of Use}}.
Objaverse~\cite{deitke2023objaverse} is available under the ODC-By v1.0 license.
For code,
ATISS~\cite{paschalidou2021atiss} is available under its NVIDIA Source Code License\footnote{\href{https://github.com/nv-tlabs/ATISS/blob/master/LICENSE}{ATISS NVIDIA Source Code License}}.
DiffuScene~\cite{tang2024diffuscene} is under its terms of use\footnote{\href{https://github.com/tangjiapeng/DiffuScene/blob/master/LICENSE}{DiffuScene Terms of Use}}.
Holodeck~\cite{yang2024holodeck}, Long-CLIP~\cite{zhang2024longclip} and Qwen2.5-VL~\cite{bai2025qwen2} are available under Apache License 2.0.
InstructScene~\cite{lin2024instructscene} and LayoutGPT~\cite{feng2024layoutgpt} are available under the MIT License.
LayoutVLM~\cite{sun2025layoutvlm} does not have a license specified in its repository.
GPT-4o~\cite{achiam2023gpt} and o4-mini~\cite{openai2025gpto4} is under the OpenAI terms of use\footnote{\url{https://openai.com/policies/terms-of-use/}}.
Our use of these datasets and code is in compliance with their respective licenses.

\section{AI Assistant Usage}
\label{sec:supp_ai_assistants_usage}

GPT-4o~\cite{achiam2023gpt} is used in this work in parts of the evaluation framework, and o4-mini~\cite{openai2025gpto4} is used in our data generation process.
For experiments involving an open-source model, we used Qwen2.5-VL~\cite{bai2025qwen2}.
We also used GitHub Copilot\footnote{\url{https://github.com/features/copilot}} and ChatGPT\footnote{\url{https://chat.openai.com/}} to assist in writing code and checking for grammatical errors and typos in the paper.

\end{document}